\documentclass{aa}
\usepackage{graphicx}
\usepackage{txfonts}
\def\tff{t_\mathrm{ff}}
\def\izw{I\,Zw\,18}
\def\sbs{SBS\,0335$-$052}
\def\Nion{\dot{N}_\mathrm{ion}}
\def\taudi{\tau_\mathrm{di}}
\def\nsii{n_{\mathrm e}([\mathrm{S}\,\mathsc{ii}])}
\def\sii{S\,{\sc ii}}
\newcommand\nion{$\dot{N}_\mathrm{ion}$}
\newcommand{\hii}{H\,\textsc{ii}}
\newcommand{\msun}{$M_\odot$}
\newcommand{\nerms}{$\langle n_{\mathrm e} \rangle$}
\newcommand{\nesii}{$n_{\mathrm e}([\mathrm{S}\,\mathsc{ii}])$}
\newcommand{\cmcubed}{\,cm$^{-3}$}
\newcommand{\hst}{\textit{HST}}
\newcommand{\ri}{$r_{\mathrm i}$}
\newcommand{\rs}{$r_{\mathrm S}$}
\newcommand{\zsun}{$Z_{\odot}$}
\newcommand{\logoh}{12$+$log(O/H)}
\begin{document}
\title{The size--density relation of extragalactic
H\,{\large II} regions}
\author{L. K. Hunt \inst{1}
\and
H. Hirashita \inst{2}
%%\fnmsep\thanks{Research Fellow of the Japan Society for the Promotion of
%%Science.}
}
\institute{INAF - Osservatorio Astrofisico di Arcetri, Largo E. Fermi, 5,
50125 Firenze, Italy\\
\email{hunt@arcetri.astro.it}
        \and
Institute of Astronomy and Astrophysics, Academia Sinica,
P.O. Box 23-141, Taipei 10617, Taiwan\\
\email{hirashita@asiaa.sinica.edu.tw}
}
\date{2 September 2009}

%\abstract{
\abstract{}{
We investigate the size--density relation in extragalactic
\hii\ regions, with the aim of understanding the role of dust
and different physical conditions in the ionized medium.
}
{First, we compiled several observational data sets 
for Galactic and extragalactic \hii\
regions and confirm that 
extragalactic \hii\ regions follow the same
size ($D$)--density ($n$) relation
%follows a relation with constant column density
as Galactic ones ($n\propto D^{-1}$), rather than a relation with
constant luminosity ($n\propto D^{-1.5}$). Motivated by
the inability of static models to explain this,
% the size--density relation of extragalactic \hii\ regions, 
we then modelled the evolution of the size--density
relation of \hii\ regions
by considering their star formation history, the effects of
dust, and  pressure-driven expansion. The results
are compared with our sample data whose size and density span 
roughly six orders of magnitude.
}
{The extragalactic samples cannot be understood as an
evolutionary sequence with a single initial condition.
Thus, the size--density relation does not result
from an evolutionary sequence of \hii\ regions but
rather reflects a sequence with different initial gas
densities (``density hierarchy'').
We also find that the size of many \hii\ regions
is limited by dust absorption of ionizing photons,
rather than consumption by ionizing neutral hydrogen.
%Such a ``dust-extinction limited'' situation could explain the
%size--density relation with constant column density
%for a given dust-to-gas ratio.
Dust extinction of
ionizing photons is particularly severe over the entire lifetime of
compact \hii\ regions with typical gas densities of
$\ga 10^3$ cm$^{-3}$. Hence,
as long as the number of ionizing photons
%ionized gas emission
%(hydrogen recombination lines, free-free radio continuum)
%in \hii\ regions 
is used to trace massive star formation, 
much star-formation activity could be missed.
Such compact dense environments, 
the ones most profoundly obscured by dust,
have properties similar to ``maximum--intensity starbursts''.
This implies that submillimeter and infrared wavelengths
may be necessary to accurately assess star formation
in these extreme conditions both locally and at high redshift.
}
{}
\keywords{dust, extinction -- galaxies: dwarf --
galaxies: evolution -- galaxies: ISM --
galaxies: star clusters -- \hii\ regions}

\titlerunning{The size--density relation of extragalactic
H\,\textsc{ii} regions}
\authorrunning{L. K. Hunt \& H. Hirashita }

\maketitle

%
%________________________________________________________________

%% the body of the paper

\section{Introduction}\label{sec:intro}

Massive stars, young star clusters, and their 
associated \hii\ regions are vital probes 
of recent star formation in nearby galaxies and the distant universe.
While Galactic \hii\ regions such as the Orion Nebula and
RCW\,49 are well studied because of their proximity,
most Galactic work is still plagued by distance uncertainties 
(e.g., Anderson \& Bania \cite{anderson09}),
although the situation is improving rapidly
(see, e.g., Hachisuka et al. \cite{hachisuka06},
Foster \& MacWilliams \cite{foster06}, Russeil et al. \cite{russeil07}).
\hii\ regions in galaxies in the Local Group can be
studied in almost as much detail as those in the Galaxy, and 
the distance estimation is much less uncertain.
However, even in Local Group galaxies, Super Star Clusters (SSCs), 
the most extreme examples of massive star formation, 
are absent.
SSCs, with $\sim10^5$--$10^6$\,\msun\
and Lyman continuum photon rates
$\Nion\sim 10^{52}$--$10^{53}$\,s$^{-1}$
enclosed in regions of $\la$10\,pc in radius,
are generally not seen in quiescent environments such
as the Milky Way, M\,31, and M\,33.
Even R136 in 30 Doradus, the most massive Local Group star cluster
(10$^{4.5}$\,\msun\ and 10$^{51.4}$\,s$^{-1}$:
Massey \& Hunter \cite{massey98}),
falls short of the typical properties of SSCs.
Hence, to study the wide range of manifestations of massive 
star formation, it is necessary to examine 
star clusters and \hii\ regions in galaxies beyond the Local Group.
Of necessity, the price to be paid is detail;
the advantage to be gained is the wide variety of 
Star-Forming (SF) complexes that can be studied.

The total mass of a cluster of massive stars
can be observationally quantified by estimating
the number of ionizing photons. A simple 
Str\"{o}mgren-sphere argument (Eq.\ \ref{eq:stromgren})
would suggest that \nion$\propto D^3n_\mathrm{e}^2$, where
\nion\ is the number of ionizing photons emitted per
unit time, $D$ is the diameter of the \hii\ region,
and $n_\mathrm{e}$ is the electron number density in the
\hii\ region. Thus, in
principle, by examining $n_\mathrm{e}$ and $D$ of
\hii\ regions, we can estimate the ionizing photon
luminosity. 
However, observationally, the situation is not straightforward.
The diameter ($D$) and the electron density
($n_\mathrm{e}$) of Galactic \hii\ regions are known to
be negatively correlated with a roughly unit slope:
$n_\mathrm{e}\propto D^{-1}$
(Garay \& Lizano \cite{garay99}; Kim \& Koo \cite{kim01};
Mart\'{\i}n-Hern\'{a}ndez et al.\ \cite{martin03};
Dopita et al.\ \cite{dopita06}). 
This would not be expected from a simple Str\"{o}mgren-sphere
argument with \nion\ constant:
$n_\mathrm{e}\propto D^{-1.5}$. 
Various possible explanations for the observed shallower slope
are presented in Mart\'{\i}n-Hern\'{a}ndez et al.\
(\cite{martin03}), including
%(see their Section 4.1 and references therein)
optical depth effects, clumpiness, dust extinction, and stellar content. 
Dust mixed in with the ionized gas has also been proposed as 
an explanation by Arthur et al.\ (\cite{arthur04}, hereafter A04)
who argue that significant absorption of ionizing photons
by dust grains in the densest \hii\ regions flattens the 
size--density relation.

Previous work on extragalactic \hii\ regions suggests that
they follow the same size--density correlation as Galactic ones.
Kennicutt (\cite{kennicutt84}) found that \hii\ regions in
spiral disks follow a $n_\mathrm{e}\propto D^{-1}$
relation, but extend the Galactic trend to lower 
densities and larger sizes.
More recently, Gilbert \& Graham (\cite{gilbert07}) examined the
SSCs in the Antennae galaxies, a prototypical starburst merger, 
and found that these also follow the $n_\mathrm{e}\propto D^{-1}$
relation.
However, the SSCs in the Antennae have densities ($\sim40-400$\cmcubed)
and sizes (25--100\,pc) that place them on a different location
in the $n_\mathrm{e}$--$D$ plane than the Kennicutt sample. 
In fact, 
Gilbert \& Graham (\cite{gilbert07}) conclude that the SSCs in 
the Antennae constitute a new class of massive \hii\ regions that 
is distinct from the Galactic and typical extragalactic population.

In this paper, we examine the size--density relation in
extragalactic \hii\ regions, and explore
several mechanisms which have been proposed to explain it,
including clumpiness, dust extinction, and stellar content.
In particular, we focus on the 
\hii\ regions in Blue Compact Dwarf galaxies (BCDs);
following Gilbert \& Graham  (\cite{gilbert07}), we shall refer to
such regions as Emission-Line Clusters (ELCs).
In fact, most of the current star formation in many BCDs occurs in
such ELCs,
rather than in the underlying diffuse component.
Previously we studied a more limited sample,
and found that the \hii\ regions in BCDs also follow the
same size--density relation namely,  $n_\mathrm{e}\propto D^{-1}$
(Hunt et al.\ \cite{hunt-cozumel}).
Here we triple the sample size used in our previous work,
and compare the BCD ELCs with other types of extragalactic \hii\
regions, such as those in spiral disks and known SSCs.

Focusing on BCDs enables us to study another aspect of the
formation of star clusters. 
Because they are generally metal poor, with oxygen abundances 
[$12+\log (\mathrm{O/H})$] ranging from 7.2 to 8.5
(Izotov, Thuan, \& Stasi{\'n}ska \cite{izotov07}),
BCDs provide a link between high-redshift metal-free primeval
galaxies and the metal-enriched SF galaxies in the 
local universe.
Because metallicity is not expected to be the only driver 
of star-formation and cluster properties, we can compare different
metallicities together with other parameters, and better disentangle
the effects of metal abundance.

In fact, the process of ionization itself in metal-poor objects
is also important in the cosmological context.  Because they are
relatively chemically unevolved, \hii\ regions in BCDs
could enable us to infer
some characteristics of the ionization processes at
high $z$.
Since the typical virial
temperature of the first-generation objects in the
Universe is $\la 10^4$ K
(e.g., Tegmark et al.\ \cite{tegmark97};
Yoshida et al.\ \cite{yoshida03}), the photoionization
which raises the gas temperature to $\sim 10^4$ K
prohibits the gas from collapsing to form stars
(Omukai \& Nishi \cite{omukai99}). The gas density
structure is modified by the pressure-driven expansion
of \hii\ regions (Kitayama et al.\ \cite{kitayama04}).
Such effects on the gas structure are of fundamental
importance for considering the subsequent star formation
and the reionization of the Universe.

The relation $n_\mathrm{e}\propto D^{-1}$ for 
Galactic and extragalactic \hii\ regions
%%commonly found for various samples as mentioned above 
can be
interpreted as a {\it constant ionized-gas column density}.
However, the physical mechanism that produces the constant
column density is not clear, nor is it clear why ELCs in external
galaxies and Galactic \hii\ regions should follow a similar relation 
with column density.
It is also not understood whether or not there are 
different classes of \hii\ regions which would occupy 
distinct zones of parameter space in the size--density 
plane, and if there are, how they might be related.

The aim of this paper is to understand the size--density
relation of extragalactic \hii\ regions, and
place it in the context of star-cluster and ELC formation.
The paper is organized as follows. First, in
Sect.\ \ref{sec:data}, we describe the observational
samples of extragalactic \hii\ regions, together
with the compilation of Galactic \hii\ regions
for comparison.
In Sect.\ \ref{sec:overall}, we examine gross trends in 
the data, which we will
later interpret in the light of our evolutionary models.
We predict the
size--density relation for {\it static} dusty \hii\ regions
by assuming a constant luminosity of the central
sources in Sect.\ \ref{sec:static}. 
Then, in Sect.\ \ref{sec:model},
we extend the model to include
pressure-driven expansion and star formation history.
Some basic results of this evolutionary model 
are given in Sect.\ \ref{sec:constraint}, and compared with
the extragalactic observational data. 
In Sect.\ \ref{sec:discussion} we discuss the models
and their implications in various contexts.
Finally we give our conclusions in Sect.\ \ref{sec:conclusion}.
We adopt a Hubble constant of
$H_0=73$ km s$^{-1}$ Mpc$^{-1}$.

\section{The data}\label{sec:data}

We have assembled several samples of \hii\ regions from the literature,
including two extragalactic and two Galactic radio data sets,
and one extragalactic optical sample. 
Another extragalactic optical sample is presented here for the
first time, with sizes measured from \hst\ images and 
electron densities derived from emission measures calculated
with published long-slit optical spectra. 

\subsection{Extragalactic radio data sets}

One extragalactic radio data set comprises those galaxies
with known ``ultra-dense \hii\ regions'' or ``radio super nebulae'' 
(see Kobulnicky \& Johnson\ \cite{kobulnickyjohnson99}, Beck et al.\ \cite{beck02}).
Many of these are found in low-luminosity low-metallicity
BCDs, but some reside
in metal-rich starbursts such as NGC\,253 and M\,82
and normal spiral disks, including NGC\,4214 and NGC\,6946.
The common feature of these multifrequency radio continuum
observations is a rising spectrum at low frequencies, and a
relatively flat spectrum at higher ones. 
This implies that the predominant emission mechanism is
thermal bremsstrahlung from ionized gas, and that the
electron density and size are such that there is a turnover
in the radio spectrum at some frequency $\nu_t$. 
This ``turnover frequency''
$\nu_t$ corresponds to an optical depth $\tau_\nu$ of unity,
and defines the frequency where the spectrum changes from 
optically thick to optically thin.
$\nu_t$ depends on the Emission Measure (EM\,=\,$\int n_e^2 d \ell$) and 
the electron temperature $T_e$.
The spectra of ``classical'' \hii\ regions such as Orion turn over at
very low frequencies, $\sim$0.3\,GHz, while the
compact dense regions in these galaxies have $\nu_t\gtrsim$5\,GHz.
All radio observations described below 
measure the high-frequency optically thin part of the
radio spectrum ($F_{\nu}\propto\nu^{-0.1}$), as well as the
transition region ($\nu\lesssim\tau_\nu$) toward lower
frequencies where the spectrum becomes optically thick.

In some cases, with high-resolution observations
(II\,Zw\,40: Beck et al. \cite{beck02};
NGC\,5253: Turner et al. \cite{turner00}), 
the radio emission is resolved at 15\,GHz.
Thus, the authors were able to estimate the size $D$ of the emitting region
by fitting a Gaussian, 
and from the observed flux, determine the EM, and
infer the rms electron density \nerms.
In other cases (NGC\,4214, NGC\,1741, Mrk\,8, Mrk\,33, VII\,Zw\,19,
Pox\,4, Tol\,35, Mrk\,1236: Beck et al. \cite{beck00}; 
NGC\,6946, NGC\,253, M\,33: Johnson et al. \cite{johnson01};
SBS\,0335-052: Hunt et al. \cite{hunt04}, Johnson et al. \cite{johnson09}),
the spatial resolution is insufficient to resolve the regions.
Then the radio spectrum can be fit by models of
homogeneous, isothermal, dust-free, ionization bounded regions of ionized
gas, to obtain the turnover frequency $\nu_t$, the EM, \nerms,
and infer the size $D$ of the region 
(e.g., Deeg et al. \cite{deeg93}; Johnson et al. \cite{johnson01};
Hunt et al. \cite{hunt04}).
Alternatively, the optically thick and optically thin regions of
the radio spectrum can be separated to constrain $\nu_t$, and
thus infer the emission measure EM, size $D$, and rms electron density \nerms\
(e.g., Gordon \cite{gordon88}; Beck et al. \cite{beck00}).
When the size is not directly measured, that is to say when the sources
are unresolved, the authors estimate sizes and densities from fits of 
multi-frequency radio continuum spectra; these are consequently not 
independent parameters, but rather negatively correlated.
Nevertheless, the logarithmic slope between density and size 
expected from this degeneracy would be $-1.5$ which is significantly
steeper than that observed (see below).
The sizes $D$ and densities \nerms\ for NGC\,5253 and He\,2$-$10
have been modeled also from radio recombination line observations, and are
consistent with those inferred from continuum fitting 
(Mohan et al. \cite{mohan01}).

There is one galaxy in our data set, I\,Zw\,18, in which 
multifrequency observations show no sign
of a rising spectrum (Hunt et al. \cite{hunt05}).
Here also the data have been fit to a model of an
homogeneous, isothermal, dust-free ionization bounded region of ionized
gas, as described above. 

We will refer to this sample of (except for I\,Zw\,18) rising-spectrum
sources as the ``radio sample'' (16 galaxies);
its mean oxygen abundance is \logoh\,=\,8.24$\pm$0.5,
or $\sim$0.22\,\zsun\footnote{In this and future
relative solar abundances, we will adopt 
the solar calibration of 
Anders \& Grevesse (\cite{anders89}).}.
When there are multiple observations for a
single object, we usually list these data as different data points.
However,
M\,33, NGC\,253, and NGC\,6946 contain several candidates for 
ultracompact \hii\ regions, 
but we show each galaxy as a single average according to
the figures in Johnson et al. (\cite{johnson01}).
The data for the radio sample are reported in Table \ref{tab:radioBCD};
all sizes have been corrected to the distance scale used here.

The other extragalactic radio sample comprises \hii\ regions in the 
Small and Large Magellanic Clouds (SMC and LMC, respectively)
(Mart\'{\i}n-Hern\'{a}ndez et al.\ \cite{martin05}), and
in the supergiant \hii\ region NGC\,604
in M\,33 (Churchwell \& Goss \cite{churchwell99}). 
As with the previous data set, these are also radio continuum
observations, but at a single frequency, 5\,GHz (SMC/LMC)
or 8.4\,GHz (NGC\,604/M\,33).
We adopt the results given in the original papers 
for sizes $D$ and densities \nerms.
The regions in these Local Group galaxies are resolved, and  
$D$ is measured from high-resolution interferometric maps
by fitting two-dimensional Gaussians, including beam deconvolution. 
The densities are inferred from the observed flux densities
by assuming that all radio emission is optically thin bremsstrahlung,
arising in a dust-free, ionization bounded, homogeneous region
with size $D$ (Mezger \& Henderson \cite{mezger67}). 
This sample will be called the ``Local-Group sample'';
with M\,33, the LMC, and the SMC, its mean oxygen abundance
is \logoh\,=\,8.41$\pm$0.3, corresponding to 0.33\,\zsun.

\subsection{Galactic radio samples}

We have also included two size--density data sets
of Galactic \hii\ regions as comparison samples. 
The first Galactic sample is taken from 
the compilation by Garay \& Lizano (\cite{garay99}), and consists of 
compact \hii\ regions observed with high angular
resolution in either the H66$\alpha$ or H76$\alpha$
lines, as well as in the radio continuum. 
The second sample is a set of (ultra) compact Galactic \hii\ regions, 
observed in the 21\,cm radio continuum by Kim \& Koo 
(\cite{kim01}).
For both samples,
we have adopted the authors' \hii\ region parameters;
the electron densities were derived assuming optically thin, 
dust-free, homogeneous, ionization-bounded nebulae 
(e.g., Mezger \& Henderson \cite{mezger67}), and
the sizes were directly measured from 
interferometric radio images by fitting Gaussians, deconvolved with the beam size.
As in the Local-Group sample,
such measurements obviate the potential density--size degeneracy with 
logarithmic slope $-1.5$ that could arise were the sizes not 
measured independently.

\subsection{\hst\ extragalactic optical sample}\label{subsec:hst}

The main optical sample includes those star-forming dwarf galaxies with 
optical spectra and usable 
high-resolution {\it Hubble Space Telescope} (\hst) archival data,
obtained either with the Advanced Camera for Surveys (ACS), 
the Wide Field Planetary Camera 2 (WFPC2),
or in the near-infrared with the NICMOS array. 
Most of the images were retrieved from the Hubble Legacy 
Archive\footnote{The Hubble Legacy Archive is a collaboration between the Space
Telescope Science Institute (STScI/NASA), the Space Telescope European
Coordinating Facility (ST-ECF/ESA) and the Canadian Astronomy Data Centre
(CADC/NRC/CSA).}.
We obtained images for 26 galaxies, but use only 23 of them; 
3 had only F160W images, and 
when images at other wavelengths were available for comparison,
the F160W images gave 
consistently larger sizes than the other wavelengths.
Virtually all of the galaxies are BCDs.

The high resolution of \hst\ is crucial in order
to resolve the ELCs in the sample objects.
Most of the galaxies are dominated by a single bright
SF complex; because we want to compare with ground-based
long-slit
optical spectra, this is the region we focus on,
rather than examining the more diffuse emission.
There were generally no H$\alpha$ images available, so we were
forced to use the continuum to determine the size of
the SF region.
Since we wanted to match the size measurement as much as possible
with the spectroscopic slit, we adopted a one-dimensional
method rather than two-dimensional models as in the radio.
For each galaxy, we measured the linear extent by fitting the
surface brightness profiles in two orthogonal cuts
with Lorentzian and Gaussian profiles. Although Lorentzians fit the 
extended wings of the profiles better than Gaussians,
they give widths that are $\sim$8\% smaller. 
Hence,
we adopted the Gaussian fits to be compatible with the analogous
fits for the radio source sizes.
The \hst\ profiles are extracted in
1-arcsec wide rectangular apertures, which match the spectroscopic
slit (i.e., with roughly the same spatial resolution as the 
H$\beta$ measurements adopted below).
The diameter of the region is defined as the geometrical mean of
the full widths at half maximum (FWHMs) of these
two orthogonal (Gaussian) profiles. 
If data are available in more than one band, we
adopt the longest-wavelength data with the highest spatial
resolution, although there is no significant
trend of sizes with filter band (except for F160W as noted above).
Despite the high spatial resolution of the \hst,
we were unable to resolve the brightest complex in
a few distant galaxies with particularly compact SF regions.
In those cases, we will be overestimating the region size,
and consequently underestimating the root-mean-square electron density;
these will be discarded in the analysis.

In general, it should be emphasized that the size measurement is a 
delicate and difficult procedure.
Many of the objects have blended regions which \hst\ did not
resolve, but which would fall within a ground-based spectroscopic slit.
Moreover, because most of the spectroscopic observations did not
give the position angle of the slit, we had to use subjective judgment
to determine the angles for our virtual cut apertures.
It is also true that ionized gas in an ELC tends to be more 
extended than the underlying nebular continuum and stellar
emission, so we are probably underestimating the size
with our method and account for this empirically (see below). 
All these considerations make the diameter determinations only good to
a factor of 2 or so, but the consistency with 
the other samples lends confidence to the procedure. 

The root-mean-square (rms) number densities \nerms\
are calculated from long-slit observations
of the optical H$\beta$ recombination line.
Following Kennicutt (\cite{kennicutt84}),
we convert the H$\beta$ surface brightness
over the spectroscopic slit area to volume emission measure. 
Hydrogen emissivities are calculated
with ionized-gas temperatures inferred from optical emission
lines, as given by published tables (see Table \ref{tab:optBCD}). 
Extinction is corrected for with published values of
$c(\textrm{H}\beta)$,
and ionized helium with a multiplicative factor of 1.08.
The continuum size of an ELC is generally 1.5 to 2 times
smaller than the ionized gas extent
(Tenorio-Tagle et al. \cite{tenorio06}, Silich et al. \cite{silich07}),
so to convert the volume emission measure to rms density \nerms,
we have adopted a region size 1.5 times as large as actually measured
in the continuum. 
To account for the larger extension of the ionized
gas relative to the continuum,
we also used this enlarged size as the ``true size'' of the \hii\
regions measured from the images.
Because of the considerable uncertainties in this entire procedure,
the densities \nerms\ are probably only good to roughly a factor of 2,
being slightly less uncertain than the diameters because of the
square-root dependence on EM.
Nevertheless,
this is similar to the uncertainty in the rising-spectrum radio
sample, and to the comparison optical sample described in Sect. \ref{sec:comparisonopt}.
The sample, which we call the ``\hst\ sample'', is given in
Table \ref{tab:optBCD}, together with
the references for the spectroscopic data. 
The mean oxygen abundance of the \hst\ sample (23 galaxies) is 
\logoh\,=\,7.89$\pm$0.35, $\sim$0.09\,\zsun.

\begin{figure}
\includegraphics[bb=18 207 568 705,width=0.5\textwidth]{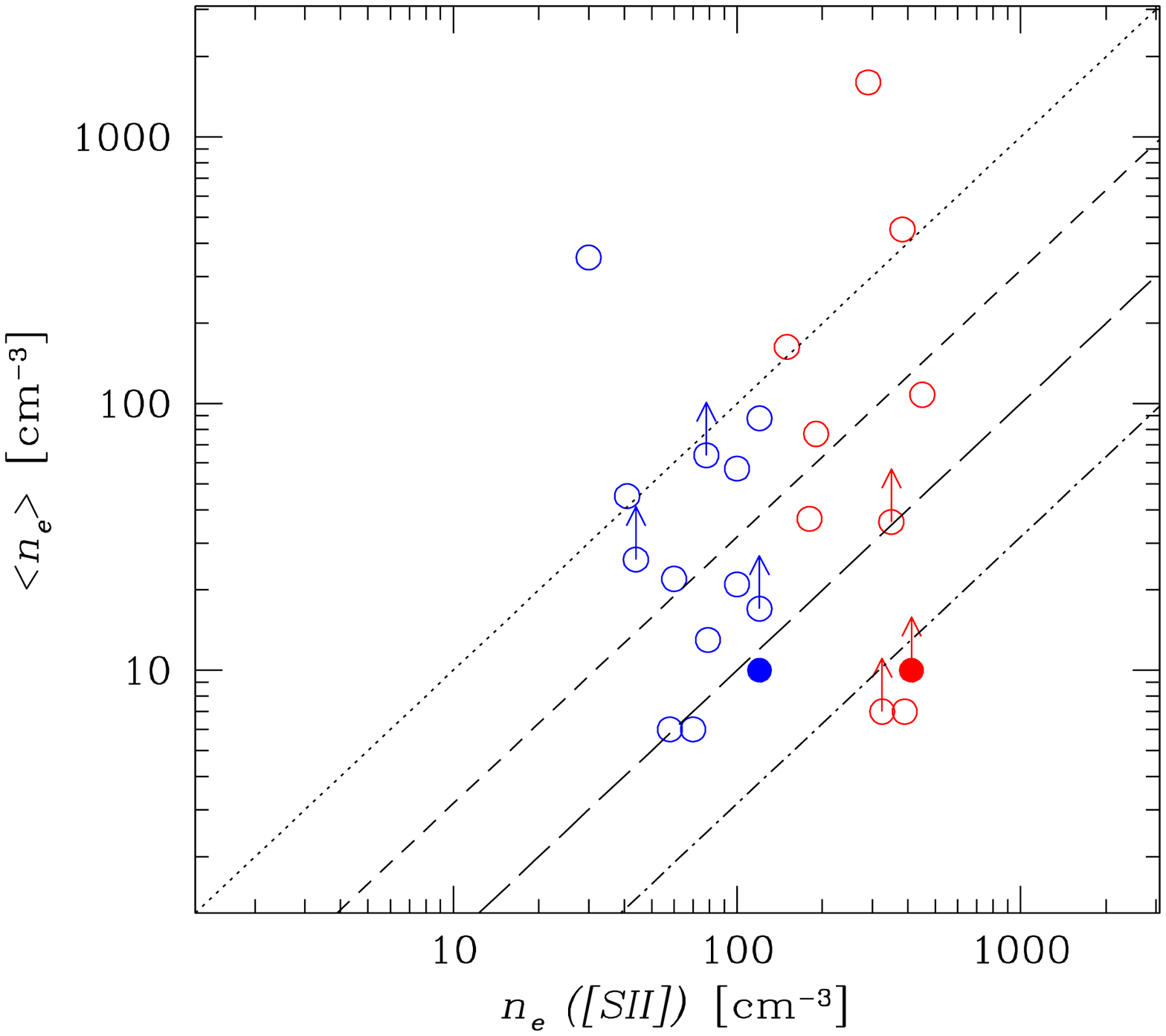}
\caption{RMS densities \nerms\ vs.
published densities inferred from the [S\,{\sc ii}] lines at 
$\lambda\lambda$6717, 6731\AA\ for the \hst\ sample.
The diagonal lines correspond to constant volume filling factors of 1
(dotted), 0.1 (short dashed), 0.01 (long dashed), and 0.001 (dot-dashed).
The filled circles correspond to \izw\ ($\nsii =120$\cmcubed),
and \sbs\  ($\nsii =412$\cmcubed), respectively. 
The red circles correspond to $\nsii >$125\cmcubed, and the blue
ones to $\nsii <$125\cmcubed.
The arrows show those galaxies with unresolved sizes, and thus
underestimated \nerms; these objects are not considered further.}
\label{fig:densities}
\end{figure}

The rms densities \nerms\ for the HST sample are plotted against the densities
determined from the
[S\,{\sc ii}] optical emission lines \nesii\ in Fig. \ref{fig:densities}.
Similarly to previous work
(Kennicutt \cite{kennicutt84}; Rozas et al. \cite{rozas98}),
the densities inferred from the [S\,{\sc ii}] line ratio \nesii\
are much higher than the rms values \nerms\ 
inferred from the emission measure.
This is because the densities measured {\it in situ} from line
ratios tend to be weighted toward
high-density high surface-brightness knots which occupy a small
fraction of the total volume
(e.g., Zaritsky et al. \cite{zaritsky94}; Kennicutt \cite{kennicutt84}).
%xxx The rms measure adopted here 
%better samples the overall ionized gas content 
%(Kennicutt \cite{kennicutt84}). 
%xxx
The difference between the two
kinds of measurements suggests that the ionized gas is clumpy,
with dense knots in a more diffuse envelope
(Kennicutt \cite{kennicutt84}; 
Zaritsky et al. \cite{zaritsky94}; Rozas et al. \cite{rozas98}). 
In a homogeneous medium with optically thin dense clumps, the
volume filling factor (FF) relates the rms density and
the sulfur-derived one: $\sqrt{\delta}\,=\,$\nerms/\nesii,
where $\delta$ signifies the filling factor.
Constant volume FFs are shown in Fig. \ref{fig:densities};
the data appear equally distributed from FFs of roughly unity to
$10^{-3}$. 
The filling factors are slightly lower, although comparable to those
in \hii\ regions in quiescent spiral disks (Kennicutt \cite{kennicutt84}).
In general, the
\hst\ sample follows the same trends in size and 
density as the other samples, and appears to be consistent with them.
The six galaxies with FWHM $<$4.5 pixels in the \hst\ images are marked with
an arrow in Fig. \ref{fig:densities}.
We are underestimating the rms densities and overestimating the
sizes for these unresolved sources,
and they are not considered in subsequent analysis. 

\subsection{Comparison extragalactic optical sample \label{sec:comparisonopt}}

The second optical sample is taken from the cornerstone study
of giant \hii\ regions in nearby spiral galaxies (Kennicutt \cite{kennicutt84}).
With ground-based photographic $\mathrm{H}\alpha$ emission-line
images, Kennicutt used a 
spherically symmetric shell model and solved the Abelian integral for 
the emission-measure profile.
Most of the profiles are monotonically decreasing with radius.
One of the ELCs measured by Kennicutt, M82-A, has also been recently measured
by another group;
the old values of (450\,pc,\,16\cmcubed; Kennicutt \cite{kennicutt84})
are now found to be (4.5\,pc,\,1800\cmcubed; Silich et al. \cite{silich07}).
Another object is in common with our sample, Mrk\,71 (NGC\,2366A=NGC\,2363):
the old values are (560\,pc,\,4\cmcubed), in contrast with our new estimate
of 14.4\,pc (this includes the doubling described above), 149\cmcubed.
The \hst\ image of Mrk\,71 gives a diameter of 7.5\,pixels ($\sim$0\farcs75), 
and clearly resolves the SF complex; this corresponds to only the brightest
portion of the region measured by Kennicutt, apparently not resolved by the
ground-based photographic images.
It is clear that the larger the region over which the density is averaged, 
the smaller the rms density.
It is noteworthy
that all these measurements, old and new, follow the same relation
between size and density, as discussed below.

\section{Overall empirical trends
\label{sec:overall}}

\begin{figure}
\centerline{
\includegraphics[bb=75 175 575 700,width=0.5\textwidth]{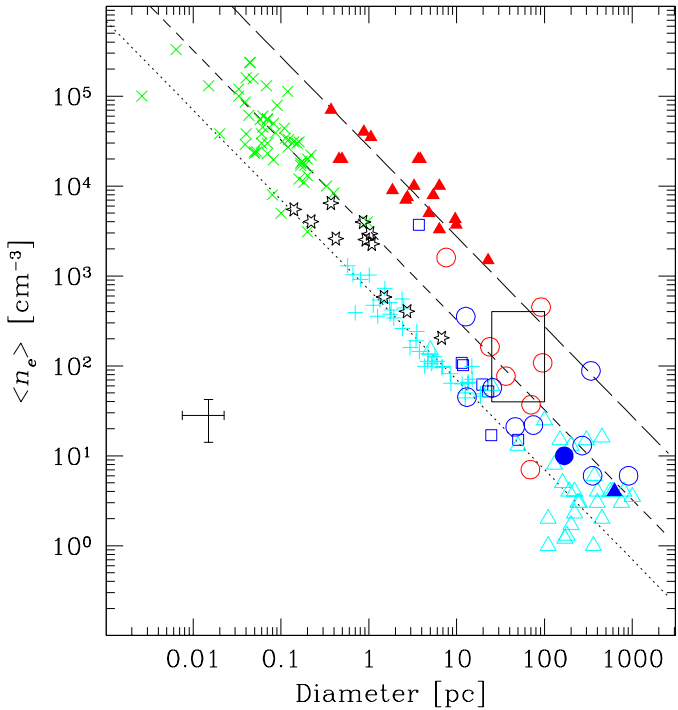}
}
\caption{Densities $n_\mathrm{e}$ vs.\ size (diameter) of the six
\hii-region samples. The extragalactic radio sample (except for M\,33
and I\,Zw\,18) is shown by filled (red) triangles; 
(the radio data for) I\,Zw\,18 by a filled blue triangle;
M\,33 by open (blue) squares;
the Garay \& Lizano (\cite{garay99}) sample by (green) $\times$;
the LMC/SMC by (black) open stars;
the Kim \& Koo (\cite{kim01}) sample by (cyan) $+$;
the Kennicutt (\cite{kennicutt84}) sample by open (cyan) triangles;
the \hst\ sample by open (red and blue) circles,
except for I\,Zw\,18 which is shown by a filled (blue) circle.
SBS\,0035$-$052 has been excluded from our \hst\ sample because it is
unresolved. 
The large rectangle shows the range of sizes and densities
for the SSCs in the Antennae galaxies (Gilbert \& Graham \cite{gilbert07}).
Root-mean-square densities \nerms\ (\cmcubed) are shown for the \hst\ sample, with 
sizes and \nerms\ calculated as described in the text.
The red circles have [S\,{\sc ii}]-inferred densities of $\gtrsim$125\cmcubed,
and blue circles $\lesssim$125\cmcubed. 
The regression lines are unit slopes (the same as the best fit, to within the
errors) with intercepts of 2.8, 3.5, and 4.4.
A factor-of-two error bar is shown in the lower left corner.
}
\label{fig:overall}
\end{figure}

The size--density relation of the samples is shown in
Fig.\ \ref{fig:overall}. All Galactic and extragalactic samples 
can be fit by $n_\mathrm{e}\propto D^{-1}$ to within the uncertainties
in the slope.
This behavior was already noted for the Galactic samples by Garay \& Lizano (\cite{garay99})
and by Kim \& Koo (\cite{kim01}).
Fig. \ref{fig:overall} shows the best-fit regressions
for the Kim \& Koo sample, the Garay \& Lizano sample, and the 
extragalactic radio sample, but with the slopes slightly tweaked to be exactly unity;
the intercepts for the three regressions ($\log n_\mathrm{e}$ [cm$^{-3}$] at
diameter = 1 pc) are 2.8, 3.5, and 4.4,
from bottom to top, respectively.
At face value, these offsets would imply that the mean column densities
in the ionized gas increase by almost two orders of magnitude,
going from compact to ultra-compact Galactic \hii\ regions, to ultra-dense
extragalactic \hii\ regions (the radio sample).
The size--density relation in all \hii\ regions,
Galactic and extragalactic, is clearly flatter
than that produced in a homogeneous Str\"{o}mgren sphere
with a constant luminosity of ionizing photons
($n_\mathrm{e}\propto D^{-1.5}$; Sect.\ \ref{sec:static}).

The trends of the different samples also suggest that metallicity
is not the key factor in \hii\ region properties.
While at approximately the same metal abundance,
some of the regions in the Local-Group sample lie closer to the
dense compact Galactic sample, while others
are coincident with the less dense Galactic sample.
Both Galactic samples are of roughly solar metallicity,
but their locations differ from one another in the size--density plane.
Lastly, the mean metallicity of the ultra-dense radio sample is
only slightly lower than solar,
but they lie far away from the locus of the (roughly solar abundance)
\hii\ regions in spiral disks.

Except for the radio sample, most of the 
extragalactic \hii\ regions follow the same size--density trend as 
the Galactic ones.  In particular, a large part of the \hst\ sample BCDs
are located at the extension of the size--density relation
of the Galactic and Local-Group samples, coincident with the \hii\
regions found in spiral disks.
The correlation of size and density in the \hst\ sample alone 
is highly significant;
with a parametric correlation coefficient $r\,=\,-0.66$, the 
(one-tailed) significance level is $\gtrsim$99.8\%.

\subsection{Emission measure and density systematics \label{sec:skeptic}}

All the densities in the size--density relation presented
here are derived from the EM of either free-free radio emission
or hydrogen recombination lines in the optical.
Because EM\,$\propto$ $D$, we might expect the 
size--density relation
to result from constant EM over a sample, combined with a constant
luminosity as in the Str\"{o}mgren argument.
The first would result in \nerms\,$\propto D^{-1/2}$, and the second
would give \nerms\,$\propto D^{-3/2}$; combining the effects would
tend to flatten the slope from the Str\"{o}mgren relation.

However, 
we can exclude this as the reason for the unit slope in the
data presented above, and in previous work by other groups.
The EMs in the Galactic radio samples vary by more than
four orders of magnitude; the same is true for the optically-inferred
EMs in the \hst\ sample. 
This would refute the hypothesis of a constant EM.
Moreover, we have verified that the densities 
inferred from the \sii\ lines for the \hst\ sample are also 
correlated with the size $D$. 
These optical measurements are independent of the EM inferred from
the hydrogen lines, and thus should provide a robust check of systematics.
Because the sulfur lines do not probe densities significantly below
$\sim$100\cmcubed, we exclude \sii\ densities 
with values $<$50\cmcubed, and find a correlation
coefficient of $r\,=\,-0.46$.
This is a weaker correlation than the one with rms densities \nerms,
but still significant at the 96\% level.
Since the \sii\ densities are independent of the emission measure,
we conclude that the size--density relation is not spuriously
induced by the method used to infer rms electron densities.

\subsection{Scale-free star formation \label{sec:scalefree}}

The power-law size--density relation of \hii\ regions
suggests that massive star formation is self-similar,
that is, there is no characteristic scale of star formation.
This scale-free nature was already noted
by Larson (\cite{larson81}) who found an approximately unit
slope between the rms H$_2$ volume density and the size of
molecular clouds: 
$\langle n(\mathrm{H}_2)\rangle\mathrm{(cm^{-3})}=3400L
\mathrm{(pc)^{-1.1}}$.
Kim \& Koo (\cite{kim01}) found a similar relation
relating Galactic
\hii\ region density and size, and argued that it reflects a variation of
the ambient density, rather than an evolutionary effect.
Indeed, the similarity of the size--density trend for molecular
clouds and \hii\ regions suggests that the star-formation processes
retain an {\it imprint} from the molecular environment
in which they take place.

This scale-free nature of \hii\ regions is also supported by the 
data presented here.
Three BCDs, He\,2$-$10, \sbs, and II\,Zw\,40, 
host both ultra-dense radio nebulae (see Table \ref{tab:radioBCD}),
and optically-visible \hii\ regions  (see Table \ref{tab:optBCD}).
All these data follow the same size--density relation, but with
different offsets.
This implies that when the same regions 
are probed with longer dust-penetrating wavelengths and higher spatial resolution, 
they turn out to be denser and smaller,
but with the same size--density relation as for the larger complexes. 
\hii\ regions and hierarchical star formation will be 
discussed further in Sect. \ref{subsec:hierarchical}.

\section{Static models}\label{sec:static}

First, we interpret the size--density relation of
the \hii\ regions compiled in Sect. \ref{sec:data} by
using simple theoretical arguments. In particular, we relate the
size and density of \hii\ regions for an
ionizing point source embedded
in a uniform and static medium with 
constant \nion. In such a situation,
the radius of the ionized region can be estimated
by the Str\"{o}mgren radius (Spitzer \cite{spitzer78}).
We also include the effect of dust extinction, which is
thought to be important in determining the
size of \hii\ regions (e.g.,
Inoue et al.\ \cite{inoue01}, A04). The models described
here are static models, and assume a  
number of ionizing photons \nion\ constant over time.

\subsection{Size--density relation of dusty static
\hii\ regions}\label{subsec:static}

Here, we estimate the radius up to which the central
source can ionize in a dusty uniform medium; this radius
is called the ionization radius, \ri.
Before estimating \ri, we define the Str\"{o}mgren
radius, $r_\mathrm{S}$, as
\begin{eqnarray}
\frac{4\pi}{3}r_\mathrm{S}^3n_\mathrm{e}n_\mathrm{H}
\alpha^{(2)}=\Nion \, ,\label{eq:ion_eq}
\end{eqnarray}
where $\alpha^{(2)}=2.6\times 10^{-13}$ cm$^3$ s$^{-1}$
is the recombination
coefficient\footnote{This value is for a gas temperature of
$T=10000$ K (Spitzer \cite{spitzer78};
Osterbrock \& Ferland \cite{osterbrock06}). If we adopt
$T=8000$ K and 20000 K, we find
$\alpha^{(2)}=3.1\times 10^{-13}$ and
% Hiro, Osterbrock (1989) is incorrect!
% see new edition with Osterbrock & Ferland for correct value.
%$2.5\times 10^{-13}$ cm$^3$ s$^{-1}$, respectively.
$1.4\times 10^{-13}$ cm$^3$ s$^{-1}$, respectively.
The temperature dependence of $\alpha^{(2)}$ does
not change our conclusions as long as we assume
a typical gas temperature $\sim 10000$ K in
\hii\ regions.}
excluding captures to the ground ($n=1$) level
(Case B),
$n_\mathrm{H}$ is the number density of hydrogen nuclei,
and $n_\mathrm{e}$ is the electron density estimated by
$n_\mathrm{e}=1.08n_\mathrm{H}$,
where the factor 1.08 indicates the correction for
helium ionization.
Numerically, the Str\"{o}mgren radius is estimated as
\begin{eqnarray}
r_\mathrm{S}=1.4\left(
\frac{n_\mathrm{H}}{10^2~\mathrm{cm}^{-3}}\right)^{-2/3}
\left(
\frac{\Nion }{10^{48}~\mathrm{s}^{-1}}\right)^{1/3}
~\mathrm{pc}\, .\label{eq:stromgren}
\end{eqnarray}

%The Str\"{o}mgren radius gives an estimate for the
%ionization radius in a
%dust-free medium. 
In the absence of dust, the ionizing radius \ri\
would be equal to \rs.
However, since a certain fraction of ionizing
photons are absorbed by dust grains, the ionization
radius \ri\ is reduced by a factor of $y_\mathrm{i}(<1)$,
which is determined by solving
(Petrosian et al.\ \cite{petrosian72};
Spitzer \cite{spitzer78}; A04)
\begin{eqnarray}
3\int_0^{y_\mathrm{i}}y^2e^{y\tau_\mathrm{Sd}}dy=1\, ,
\label{eq:yi_int}
\end{eqnarray}
where $\tau_\mathrm{Sd}$ is the dust optical depth for the
ionizing photons over a path length equal to the
Str\"{o}mgren radius. 
This integral can be expanded: %%, and solved for $y_\mathrm{i}$:
\begin{eqnarray}
1 & = & \frac{3y_i^2}{\tau_{\mathrm{Sd}}} e^{\tau_{\mathrm{Sd}} y_{\mathrm i}}
- \frac{6y_{\mathrm i}}{\tau_{\mathrm{Sd}}^2} e^{\tau_{\mathrm{Sd}} y_{\mathrm i}}
+ \frac{6}{\tau_{\mathrm{Sd}}^3} \left( e^{\tau_{\mathrm{Sd}} y_{\mathrm i}} - 1 \right)
\label{eq:yi}
\end{eqnarray}
In the optically thin limit ($\tau_\mathrm{Sd}\ll 1$), Eq.\ (\ref{eq:yi})
becomes $y_\mathrm{i}\simeq 1-\tau_\mathrm{Sd}/4$. Thus,
we confirm that $y_\mathrm{i}\to 1$ (i.e.,
$r_\mathrm{i}\to r_\mathrm{S}$) as $\tau_\mathrm{Sd}\to 0$.

Once we obtain $y_\mathrm{i}$ for a value of $\tau_\mathrm{Sd}$
(given in Sect.\ \ref{subsec:taud}),
the ionization radius can be estimated as
\begin{eqnarray}
r_\mathrm{i}=y_\mathrm{i}r_\mathrm{S}\, .\label{eq:ri}
\end{eqnarray}
We also define the optical depth over the ionization
radius as
\begin{eqnarray}
\tau_\mathrm{di}=y_\mathrm{i}\tau_\mathrm{Sd}\, .
\label{eq:taudi}
\end{eqnarray}

\subsection{Dust optical depth $\tau_\mathrm{Sd}$}
\label{subsec:taud}

In the above, we have left $\tau_\mathrm{Sd}$ undetermined.
Hirashita et al.\ (\cite{hirashita01}, 
see Appendix \ref{app:analytic} for an alternative derivation) 
estimate it under a uniform dust-to-gas mass ratio $\mathcal{D}$ as
\begin{eqnarray}
\tau_\mathrm{Sd}  =  0.87\left(\frac{\mathcal{D}}{6~10^{-3}}
\right)\left(\frac{n_\mathrm{H}}{10^2~\mathrm{cm}^{-3}}
\right)^{1/3}\left(
\frac{\Nion }{10^{48}~\mathrm{s}^{-1}}
\right)^{1/3}\, .\label{eq:tauSd}
\end{eqnarray}
The dust-to-gas ratio of the solar
neighborhood is assumed to be $\mathcal{D}_\odot =6~10^{-3}$
(Spitzer \cite{spitzer78}). We adopt various
constant values for $\mathcal{D}$ and do not follow
the time evolution of $\mathcal{D}$ in order to
avoid uncertainty in the parameters concerning
the chemical evolution models.
%%Throughout this paper, the dust-to-gas ratio
%%is uniform and is the same between neutral and
%%ionized media. 

From the equations in Sects.\ \ref{subsec:static}
and \ref{subsec:taud}, we can infer the qualitative behavior of
$r_\mathrm{i}$ as a function of \nion.
As \nion\ increases, $r_\mathrm{S}$ and $\tau_\mathrm{Sd}$ increase
(Eqs.\ \ref{eq:stromgren} and \ref{eq:tauSd})
with $\propto\Nion^{1/3}$.
Inspection of Eq.\ (\ref{eq:yi}) suggests
that if $\tau_\mathrm{Sd}\ga 1$, $y_\mathrm{i}$ drops in a
very sensitive manner with
an increase of $\tau_\mathrm{Sd}$ because of the
exponential dependence. 
Thus, if
$\tau_\mathrm{Sd}\gg 1$, $r_\mathrm{i}$ increases only
slightly with \nion\  because $y_\mathrm{i}$ decreases
significantly.
On the contrary,
$y_\mathrm{i}\sim 1$ if $\tau_\mathrm{Sd}<1$, since the
dust extinction is not severe. In this case,
$r_\mathrm{i}$ is roughly proportional to \nion$^{1/3}$.

\subsection{Dust-to-gas ratio, metallicity, and filling factor 
\label{subsec:kappa}}

It is important to realize that $\mathcal{D}$ in
$\tau_\mathrm{Sd}$ (Eq.\ \ref{eq:tauSd}) should not be interpreted
strictly as a dust-to-gas ratio dependent on metallicity.
As implied by Fig. \ref{fig:densities}, the ionized gas must be
clumpy, with concentrations of dense gas embedded in a more
tenuous medium (Sect.\ \ref{subsec:hst}).
Following Kennicutt (\cite{kennicutt84}) and 
Osterbrock \& Flather (\cite{osterbrock59}),
we assume
that the emission (and mass) of the ionized gas is dominated by the
dense clumps. 
In this case, the much less dense inter-clump region would
provide a negligible contribution to the gas emission and mass.
Implicit in this assumption is the optically-thin nature of the
clumps (c.f., Giammanco et al. \cite{giammanco04}).
Therefore, the rms formulation of our models together with the assumption
of a uniform distribution would dictate the
introduction of a gas volume filling factor $\delta$.
Eq.\ (\ref{eq:tauSd}) then becomes:
\begin{eqnarray}
\tau_\mathrm{Sd}  =  0.87\left(\frac{\mathcal{D}}{6~10^{-3}}
\right)
\left(\delta^{-1/6} \right)
\left(\frac{n_\mathrm{H}}{10^2~\mathrm{cm}^{-3}}
\right)^{1/3}\left(
\frac{\Nion }{10^{48}~\mathrm{s}^{-1}}
\right)^{1/3} ,\label{eq:tauSdfill}
\end{eqnarray}
%The factor $\delta^{-1/6}$ comes from the fact that 
Because only a fraction
of the volume $\delta$ ($\delta^{1/3}$ of the path length) is
occupied by the dense gas,
a factor of $\delta^{-1/6}$ must be introduced:
$\tau_{\rm Sd}\propto n_{\rm H}^\mathrm{dense}
(\delta^{1/3}r_\mathrm{S})
\propto n_\mathrm{H}^\mathrm{rms}\delta^{-1/2}\delta^{1/3}r_\mathrm{S}
\propto n_\mathrm{H}^\mathrm{rms}r_\mathrm{S}\delta^{-1/6}$
(``rms'' and ``dense'' distinguish between the rms density
and the density in the clumps).
The rather mild dependence on $\delta$ means that a filling factor of
0.001 would cause only an increase of a factor of $3$ in
$\tau_{\rm Sd}$. 

For convenience, we define the dust-to-gas ratio $+$ filling factor
normalized to the solar neighborhood value, $\kappa$,
as
\begin{eqnarray}
\kappa\equiv\delta^{-1/6}\mathcal{D}/\mathcal{D}_\odot\, .
\end{eqnarray}
In our models, we have assumed a uniform dust
distribution within the \hii\ region, since we have
maintained a constant dust-to-gas ratio $+$ filling factor. 
%(dust-to-gas relative filling factor). 
However, because of stellar winds 
or grain evaporation,
the central cavity surrounding the star
cluster could be devoid of dust
(e.g., Natta \& Panagia \cite{natta76}; Inoue \cite{inoue02}).
In Galactic \hii\ regions excited by single stars,
volumes with sizes $\sim0.5-0.8$ \rs\ can be dust-free, which would
imply dust filling factors as low as $\sim$40\%.
Because of grain destruction effects and strong stellar winds,
the dust-free volumes could be even larger in extragalactic \hii\ regions
ionized by massive compact star clusters with $10^3-10^4$ O\ stars.
In this case, we can still use our formulation,
but should interpret the ``effective'' dust-to-gas ratio $\kappa$
as being small because of the central cavity. Thus, as mentioned above,
the interpretation of small $\kappa$ is not unique: it could be due 
either to a low dust content, a relatively high gas filling factor, 
or to a non-uniform dust distribution
caused perhaps by a dust cavity.
It is likely that all three mechanisms contribute to the meaning of 
$\kappa$ in our formalism.

\subsection{Results}\label{subsec:static_result}

\begin{figure*}
\sidecaption
\hbox{
\includegraphics[height=6cm]{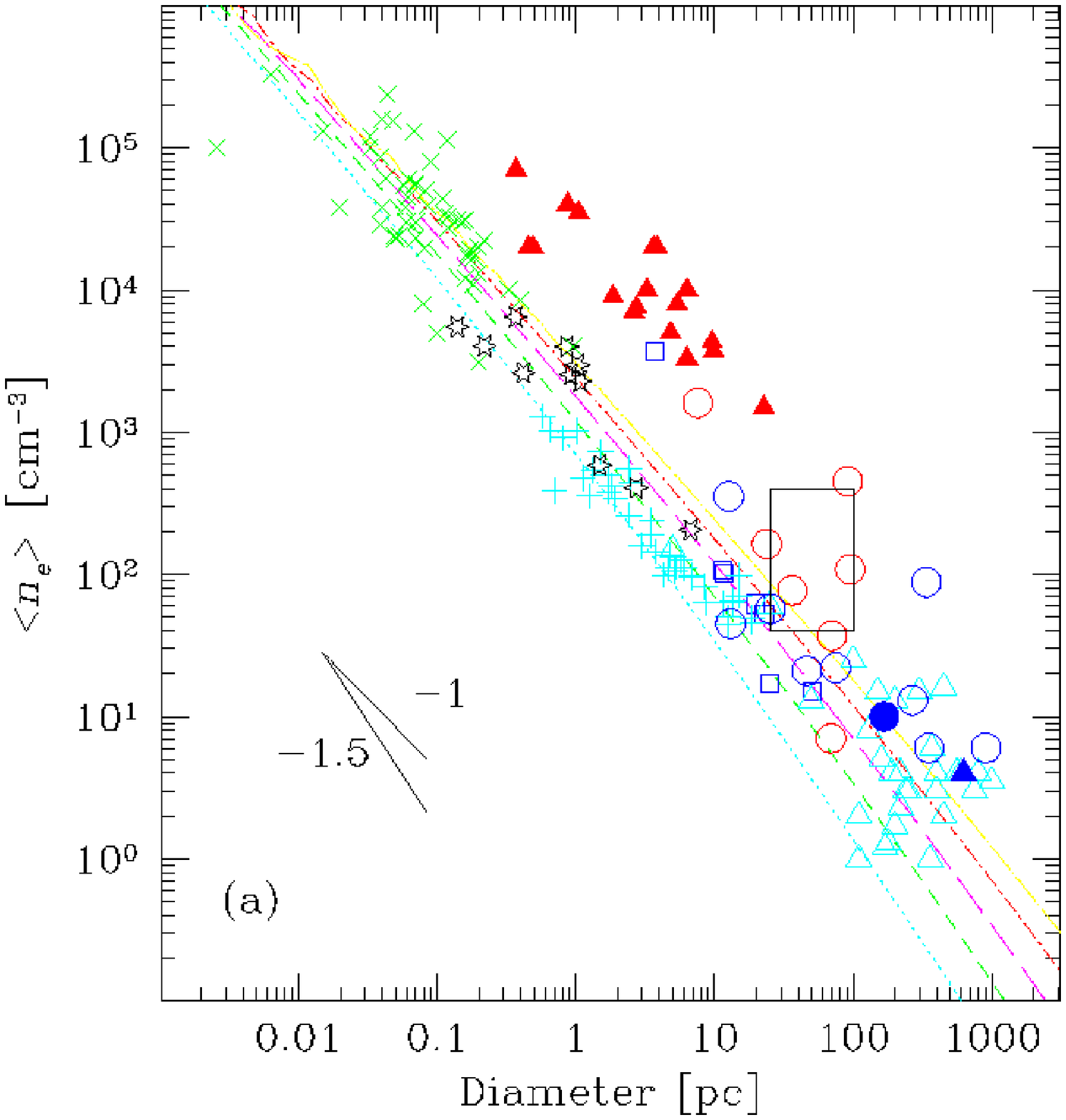}
\hspace {-0.2cm}
\includegraphics[height=6cm]{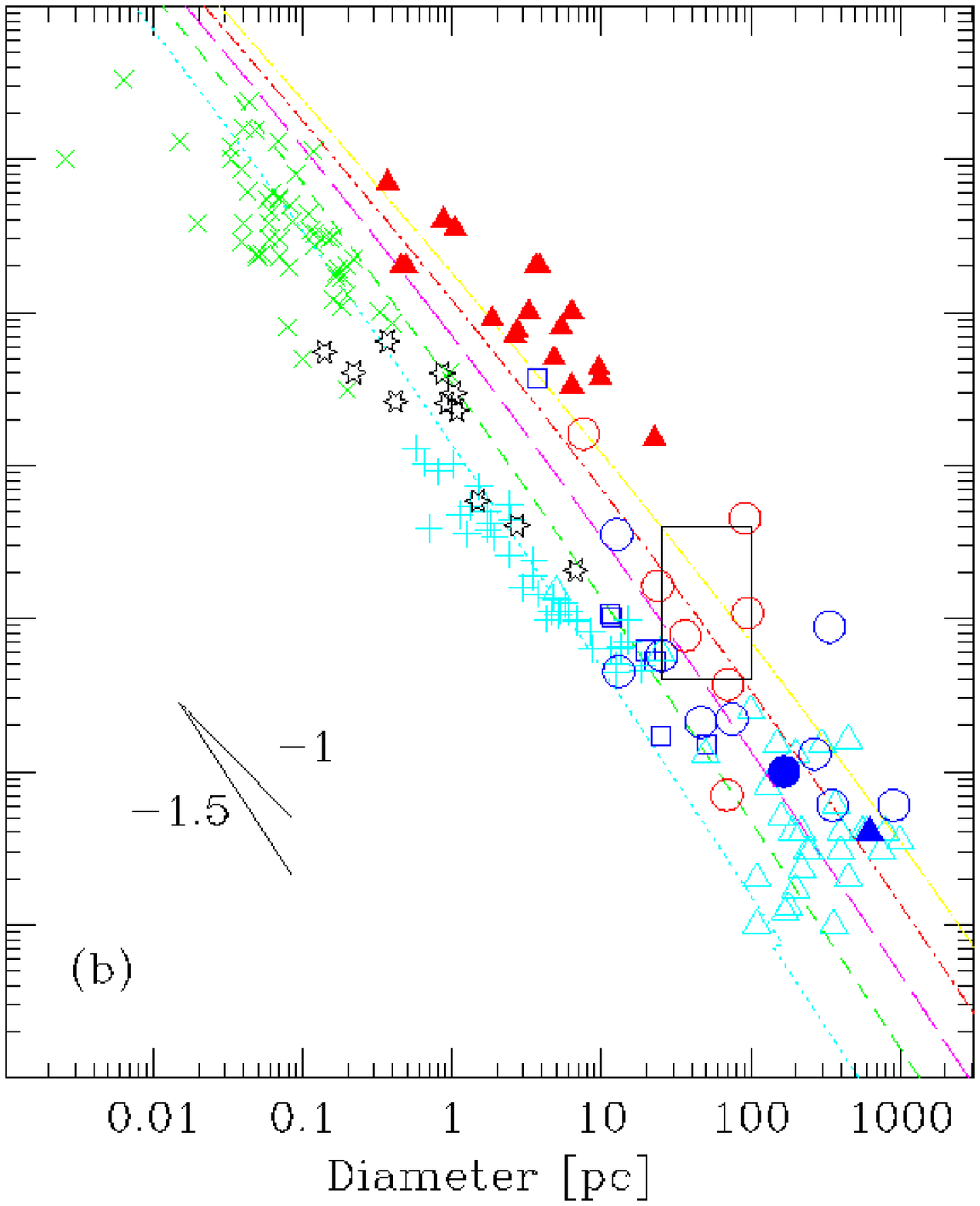}
}
\caption{Relation between the rms electron number density
\nerms\ and the ionization diameter $D_\mathrm{i}$ for
various samples. 
Symbols are as in Fig. \ref{fig:overall}.
In addition to the observational data, some
theoretical predictions for the static models are shown
in each panel:
(a) Left panel: the dotted (cyan), short-dashed (green), 
long-dashed (magenta), dot-dashed (red), and 
dot-long-dashed (yellow) lines present the results with
$N_\mathrm{ion}=10^{49}$, $10^{50}$,
$10^{51}$, $10^{52}$, and $10^{53}$ s$^{-1}$, respectively. 
A Galactic dust-to-gas ratio $+$ filling factor ($\kappa =1$) is assumed.
(b) Right panel: same as Panel (a) but with $\kappa =0.1$. 
The regression slopes
corresponding to $n_\mathrm{e}\propto D_\mathrm{i}^{-1}$
and $n_\mathrm{e}\propto D_\mathrm{i}^{-1.5}$ are also presented
in the lower-left corner in each panel.}
\label{fig:static}
\end{figure*}

In Fig.\ \ref{fig:static}, the results of the static
models are plotted over the observational samples
for various \nion.
Hirashita et al.\ (\cite{hirashita01}) suggest that
$\Nion=3\times 10^{49}$ s$^{-1}$ on average for
Galactic \hii\ regions. Indeed
$\Nion\sim 10^{49}$--$10^{50}$ s$^{-1}$
is consistent
with the Galactic \hii\ region sample as shown in
Fig.\ \ref{fig:static}a. In
Fig.\ \ref{fig:static}a, the dust-to-gas ratio is
assumed to be Galactic
($\kappa =1$), while in Fig.\ \ref{fig:static}b,
$\kappa =0.1$ to take into
account the relatively low-metallicity of the 
\hst\ sample
(Table \ref{tab:optBCD}). As
shown in Fig.\ \ref{fig:static}a,
the size--density relation is relatively insensitive to
the change of \nion. The reason for this
is described in the last paragraph in
Sect.\ \ref{subsec:taud}; since $\tau_\mathrm{Sd}\gg 1$,
the increase of $r_\mathrm{S}$ with \nion\ is
compensated by the decrease of $y_\mathrm{i}$, and as
a result, $r_\mathrm{i}$ increases only slightly.

Because of this weak dependence of $r_\mathrm{i}$
on \nion,
it is extremely difficult to explain the
data of some BCDs, unless we assume an extremely large \nion.
However, the size--density relation of the extragalactic
sample, especially that of the
BCDs, is readily explained if we assume a lower
dust-to-gas ratio typical of the BCD sample
($\kappa =0.1$), as shown in Fig. \ref{fig:static}b. 
For this value of dust-to-gas ratio,
$\tau_\mathrm{Sd}$ is typically $\la 1$, and
$r_\mathrm{i}$ increases almost in proportion to
\nion$^{1/3}$.

%If dust extinction is absent, the size--density
%relation under a constant \nion\ follows
%$n\propto D^{-3/2}$. In the dense extremes,
%the size--density relation becomes shallower (A04),
%explaining the Galactic compact \hii\ region sample.
% this is false... A04 use dust to explain the shallowness.
%However, the large less dense Galactic \hii\ regions also have
%a shallower slope than $n\propto D^{-3/2}$ in the
%size--density relation. Moreover, the extragalactic
%samples also have size--density relations described
%as $n\propto D^{-1}$ rather than $n\propto D^{-3/2}$,
%and thus cannot be explained by constant \nion.

\begin{figure}
\includegraphics[bb=18 216 556 648,width=0.5\textwidth]{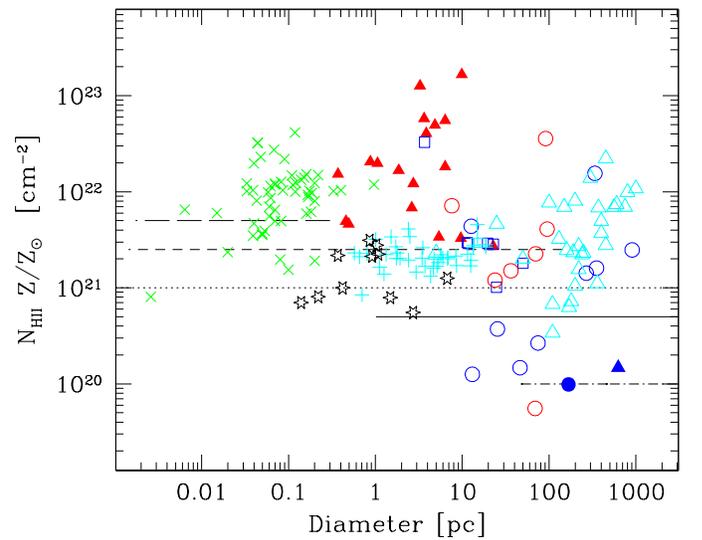}
\caption{Ionized gas column densities (cm$^{-2}$) vs. diameter (pc).
The ionized gas column densities have been multiplied by their
oxygen abundance relative to solar, assuming the 
Anders \& Grevesse (\cite{anders89})
calibration.
Constant $\taudi$ are plotted as horizontal lines:
$\taudi =$ 0.2 (dot-dashed), 1 (solid), 2 (dotted), 5 (short-dashed), and
10 (long-dashed).
Symbols are as in Fig. \ref{fig:overall}.
}
\label{fig:coldens}
\end{figure}

Because the size--density relation implies a constant
ionized gas column density, as outlined in the Introduction,
we might instead expect that the data would be consistent with
constant $\taudi$, the dust optical depth within the
ionization radius.
To test this, because the dust content is expected to 
decrease with decreasing metallicity,
we should correct the inferred ionized gas column densities
for the different metal abundances of the samples.
Fig.\ \ref{fig:coldens} shows the ionized gas column densities
multiplied by their oxygen abundance relative to solar plotted
against region diameter.
This correction assumes that $\kappa\propto Z$/\zsun,
where $Z$ is metallicity or oxygen abundance.
The horizontal lines\footnote{In our static model,
the corrected column densities correspond to
$4.9\ 10^{20}\taudi$\,cm$^{-2}$,
assuming a Galactic dust-to-gas ratio of $6\times 10^{-3}$ (see Appendix \ref{app:analytic}).}
correspond to constant $\taudi$ 
of 0.2, %(dot-dashed)
1, %(solid), 
2, %(dotted)
5, %(short-dashed), 
and 10. %(long-dashed).
For these calculations of $\taudi$, \nion\
ranged from $10^{48}$ to $10^{53}$\,s$^{-1}$,
and electron densities \nerms\ from $10^{-2}$ to $10^{6}$\cmcubed.
A large optical depth of $\taudi\sim 10$ is only possible for
large ionization radii achieved with
high values of \nion\ ($\sim 10^{52} $--$10^{53}$\,s$^{-1}$)
and high densities
($\langle n_\mathrm{e}\rangle\sim10^4$--$10^6$\cmcubed). 
This is why the $\taudi=10$ horizontal line is shorter than the others. 
Conversely, the low optical depth $\taudi=0.2$ value can only be achieved for 
low values of \nion\ and low densities,
($\Nion\sim 10^{48}$--$10^{50}$\,s$^{-1}$,
$\langle n_\mathrm{e}\rangle\sim10^{-2}$--1\cmcubed. 
This is why the $\taudi=0.2$ line lies at large diameters.

It is clear from Fig. \ref{fig:coldens} that the corrected
data are also inconsistent with constant $\taudi$.
Moreover, a linear correction of dust column density (or optical depth
$\taudi$)
for metallicity does not align all the samples. Hence, such a correction
is apparently either insufficient or wrong, when considering the dust content of
low-metallicity SF regions, at least in the context of these
simple models.

The above implies that the size--density relation is
not a sequence of either \nion\ or $\taudi$.
Rather it is probable that the relation should be
considered with variable \nion. This means that
we should assess how \nion\ evolves if we
insist that the entire sample should be explained
by a single ``unified'' sequence.
Thus, in the following we include the time variation of
\nion\ into our model, and calculate the consequent variations
in density and size.

\section{Evolutionary models}\label{sec:model}

The treatment used for the evolution of an \hii\ region
is based on our previous paper,
Hirashita \& Hunt (\cite{hirashita06}, hereafter
HH06), where we treat the evolution of the number of
ionizing photons emitted per unit time
(\nion) under a given star formation
history. We extend our models to include the effect of
grains according to A04. 
Indeed, as shown below, dust absorption of ionizing photons
significantly reduces the size of \hii\ regions especially
for compact \hii\ regions (Sect.\ \ref{subsec:extinc}).
We describe our models in the following.

\subsection{Star formation rate}\label{subsec:sfr}

We assume a spherically symmetric uniform SF
region with initial hydrogen number density,
$n_\mathrm{H0}$, and available gas mass for the star
formation, $M_\mathrm{gas}$. Following HH06, we relate the
star formation rate (SFR) with the free-fall time of
gas. The free-fall time, $\tff$, is evaluated as
\begin{eqnarray}
\tff =
\frac{4.35\times 10^7}{\sqrt{\,\mathstrut n_\mathrm{H0}\,}}
~\mathrm{yr}\, .
\label{eq:freefall}
\end{eqnarray}

Then, the SFR, $\psi (t)$, is estimated with the gas mass
divided by the free-fall time as
\begin{eqnarray}
\psi (t) & = &
\frac{\epsilon_\mathrm{SF}M_\mathrm{gas}}{\tff}f(t)\nonumber\\
& \simeq & 0.230\left(
\frac{\epsilon_\mathrm{SF}}{0.1}\right)
\left(\frac{M_\mathrm{gas}}{10^7~M_\odot}\right)\left(
\frac{n_\mathrm{H0}}{100~\mathrm{cm}^{-3}}
\right)^{1/2}f(t)~M_\odot~\mathrm{yr}^{-1},
\label{eq:sfr}
\end{eqnarray}
where $\epsilon_\mathrm{SF}$ is the star formation efficiency
assumed to be 0.1 throughout this paper (HH06),
$M_\mathrm{gas}$ is the total gas mass available in the
SF region, and the dimensionless
function $f(t)$ specifies the
functional form of the SFR.
We set the zero point of time $t$ at the onset of star
formation in the SF
region.
We investigate the following functional form for $f(t)$:
\begin{eqnarray}
f(t)= %%f_\mathrm{c}(t)\equiv
\left\{
\begin{array}{@{\,}ll@{\,}}
%\exp (-\epsilon_\mathrm{SF}t/\tff ) &
\exp (-t/t_{\mathrm g} ) &
\mbox{if $t\geq 0$,} \\
0 & \mbox{if $t<0$,}\label{eq:continuous}
\end{array}
\right.
\end{eqnarray}
where we define the gas consumption timescale
$t_\mathrm{g}$ as
\begin{eqnarray}
t_\mathrm{g}=\tff /\epsilon_\mathrm{SF}\, .
\end{eqnarray}
We will refer to this form of the SFR as ``exponentially decaying''.
In order to examine the effects of continuous
(not decaying) input of
ionizing photons, we also adopt another functional
form for the SFR called ``constant SFR'':
\begin{eqnarray}
f(t)= %%f_\mathrm{c}(t)\equiv
\left\{
\begin{array}{@{\,}ll@{\,}}
1 &
\mbox{if $t\geq 0$,} \\
0 & \mbox{if $t<0$ or $t>t_\mathrm{g}$,}
\label{eq:instantaneous}
\end{array}
\right.
\end{eqnarray}
We stop the calculation at $t=t_\mathrm{g}$ in
the constant SFR, since
after this time the effect of gas consumption should
modify the
density structure around the ionizing source.

%%can be related
%%to the radius and the total gas mass of the
%%star-forming region as (H04)
%%\begin{eqnarray}
%%n_\mathrm{H0}\simeq 69\left(\frac{r_\mathrm{SF}}{100~\mathrm{pc}}
%%\right)^{-3}\left(\frac{M_\mathrm{gas}}{10^7~M_\odot}
%%\right)~\mathrm{cm}^{-3}\, .\label{eq:density}
%%\end{eqnarray}

\subsection{Evolution of ionizing photon luminosity}

The evolution of the number of ionizing photons emitted
per unit time, \nion, is calculated by
(HH06)
\begin{eqnarray}
\Nion (t)=\int_0^\infty\mathrm{d}m\int_0^{\tau_m}
\mathrm{d}t'\, Q(m)\,\phi (m)\,\psi (t-t')\, ,
\end{eqnarray}
where $\tau_m$ is the lifetime of a star with mass $m$,
$Q(m)$ is the number of ionizing photons emitted by a
star with mass $m$ per unit time, and
$\phi (m)$ is the initial mass function (IMF).
We assume a Salpeter IMF $\phi (m)\propto m^{-2.35}$
(Salpeter \cite{salpeter55}) with stellar mass range of
0.1--100 $M_\odot$. The functional form of the SFR
$\psi (t)$ is specified in Sect.\ \ref{subsec:sfr}.
We use the fitting formulae of Schaerer (\cite{schaerer02})
for $\tau_m$ and $Q(m)$.
We adopt $Z=0$ (zero metallicity) for the stellar
properties following HH06. If we adopt solar metallicity,
\nion\ decreases by a factor of $\sim 2$ for a given stellar mass, 
but the decrease of
\nion\ for metal-rich stars can be compensated if
we assume $\sim 2$ times larger SFR.
In fact, after taking into account the different SFR and
initial gas mass, our results are comparable to those in Starburst\,99
(Leitherer et al. \cite{sb99}).

\subsection{Radius of the ionized region}\label{subsec:ri}

In order to treat the range of evolution variations from
deeply embedded \hii\ regions to normal \hii\ regions,
it is crucial to include pressure-driven expansion of
\hii\ regions. We adopt a simple analytical 
approximation based on HH06. Here, we newly include
the effect of dust, since A04 have shown that the
dust extinction significantly reduces the radius of
compact \hii\ region.

We divide the growth of an \hii\ region into two
stages: the first stage is the growth of ionizing front
due to the increase of ionizing photons, and the second
is the pressure-driven expansion of ionized gas. The
expansion speed of ionizing front in the first stage
is simply estimated by the increasing rate of the
ionization radius (Eq.\ \ref{eq:ri}). We denote the
Str\"{o}mgren radius under the initial density as
$r_\mathrm{S0}$, which is estimated by using
Eq.\ (\ref{eq:ion_eq}) with $n_\mathrm{H}=n_\mathrm{H0}$.
Because of dust extinction, the ionization radius
$r_\mathrm{i}$ is reduced by a factor of $y_\mathrm{i}$,
which is determined by solving Eq.~(\ref{eq:yi}),
where $\tau_\mathrm{Sd}$ is evaluated by
Eq.~(\ref{eq:tauSd}) with a given $\mathcal{D}$ (or
$\kappa$). The ionization radius $r_\mathrm{i}$ and
$\taudi$ are
estimated by Eqs.\ (\ref{eq:ri}) and (\ref{eq:taudi}),
respectively.

Initially, the increase of $r_\mathrm{i}$ is caused by
the accumulation of ionizing stars. Roughly speaking,
as long as $\dot{r}_\mathrm{i}$ (the
increase rate of the ionization radius) is
larger than the sound speed of ionized gas,
$C_\mathrm{II}$ (we assume $C_\mathrm{II}=10$ km s$^{-1}$
in this paper), the ionizing front propagates before
the system responds hydrodynamically. Therefore,
we neglect the hydrodynamical expansion if
$\dot{r}_\mathrm{i}>C_\mathrm{II}$, and adopt the fixed
density $n_\mathrm{H0}$.

Once $\dot{r}_\mathrm{i}<C_\mathrm{II}$ is satisfied,
pressure-driven expansion is treated. Since the density
evolves, we calculate the Str\"{o}mgren radius $r_\mathrm{S}$
under the current density $n_\mathrm{H}$ by using
Eq.\ (\ref{eq:ion_eq}).
In this situation, the growth of the ionizing region
is governed by the pressure of ionized gas and the
luminosity change of the central stars has only a minor
effect. Therefore, the following equation derived by
assuming a constant luminosity (A04)
approximately holds:
\begin{eqnarray}
\left(\frac{\dot{r_\mathrm{i}}}{C_\mathrm{II}}\right)^2
=\frac{\rho_\mathrm{II}}{\rho_\mathrm{I}}\left[1-
\frac{\rho_\mathrm{II}}{\rho_\mathrm{I}}g(\tau_\mathrm{di})^{-2}
\right]^{-1}\, ,
\label{eq:expand1}
\end{eqnarray}
where $\rho_\mathrm{I}$ and $\rho_\mathrm{II}$ are the
gas densities outside and inside of the ionized region,
respectively, and the function $g(\tau_\mathrm{di})$ is
defined as
\begin{eqnarray}
g(\tau_\mathrm{di})=\frac{3}{\tau_\mathrm{di}^3}(2
e^{-\tau_\mathrm{di}}-2+2\tau_\mathrm{di}-\tau_\mathrm{di}^2+
\tau_\mathrm{di}^3)\, .
\end{eqnarray}
%%Here, $\tau_\mathrm{di}$ is evaluated with the
%%current number density in the ionized region by
%%using Eqs.\ (\ref{eq:yi})--(\ref{eq:taudi})
%%($r_\mathrm{i0}$ is substituted with $r_i$).
The density
ratio $\rho_\mathrm{II}/\rho_\mathrm{I}$ is estimated by (A04)
\begin{eqnarray}
\frac{\rho_\mathrm{II}}{\rho_\mathrm{I}} & \simeq &
\left(\frac{r_\mathrm{i0}}{r_\mathrm{i}}
\right)^{3/2}\left(\frac{\tau_\mathrm{di}}{\tau_0}
\right)^{3/2}\left[
\frac{-2+e^{\tau_0}(2-2\tau_0+\tau_0^2)}
{-2+e^{\tau_\mathrm{di}}
(2-2\tau_\mathrm{di}+\tau_\mathrm{di}^2)}
\right]\, ,\label{eq:expand2}
\end{eqnarray}
where $r_\mathrm{i0}$ is evaluated by the Str\"{o}mgren
radius under the current ionizing photon luminosity and
the initial gas density $n_\mathrm{H0}$, and $\tau_0$
is the dust optical depth through $r_\mathrm{i0}$ under
density $\rho_\mathrm{I}$. Since $n_\mathrm{H}$ is
defined in the ionized
region, we relate it with $\rho_\mathrm{II}$ as
\begin{eqnarray}
\rho_\mathrm{II}=\mu m_\mathrm{H}n_\mathrm{H}\, ,
\end{eqnarray}
where $\mu m_\mathrm{H}$ is the gas mass per hydrogen
atom. The density in the neutral region is assumed
to be constant:
\begin{eqnarray}
\rho_\mathrm{I}=\mu m_\mathrm{H}n_\mathrm{H0}\, .
\end{eqnarray}
Eqs.\ (\ref{eq:expand1}) and (\ref{eq:expand2})
are numerically integrated to obtain $r_\mathrm{i}$ as
a function of $t$.

The above pressure-driven expansion is treated as long
as $\mathrm{d}(y_\mathrm{i}r_\mathrm{S})/\mathrm{d}t>0$.
When the SFR declines significantly, $r_\mathrm{S}$
begins to decrease. Thus,
$\mathrm{d}(y_\mathrm{i}r_\mathrm{S})/\mathrm{d}t<0$ may
be satisfied at a certain time. When
$\mathrm{d}(y_\mathrm{i}r_\mathrm{S})/\mathrm{d}t<0$,
$r_\mathrm{i}=y_\mathrm{i}r_\mathrm{S}$ is adopted with
fixed $n_\mathrm{H}$; that is, we finish treating the
dynamical expansion.

\section{Initial conditions and results }
\label{sec:constraint}

\begin{figure*}
\sidecaption
\hbox{
\includegraphics[height=4.0cm]{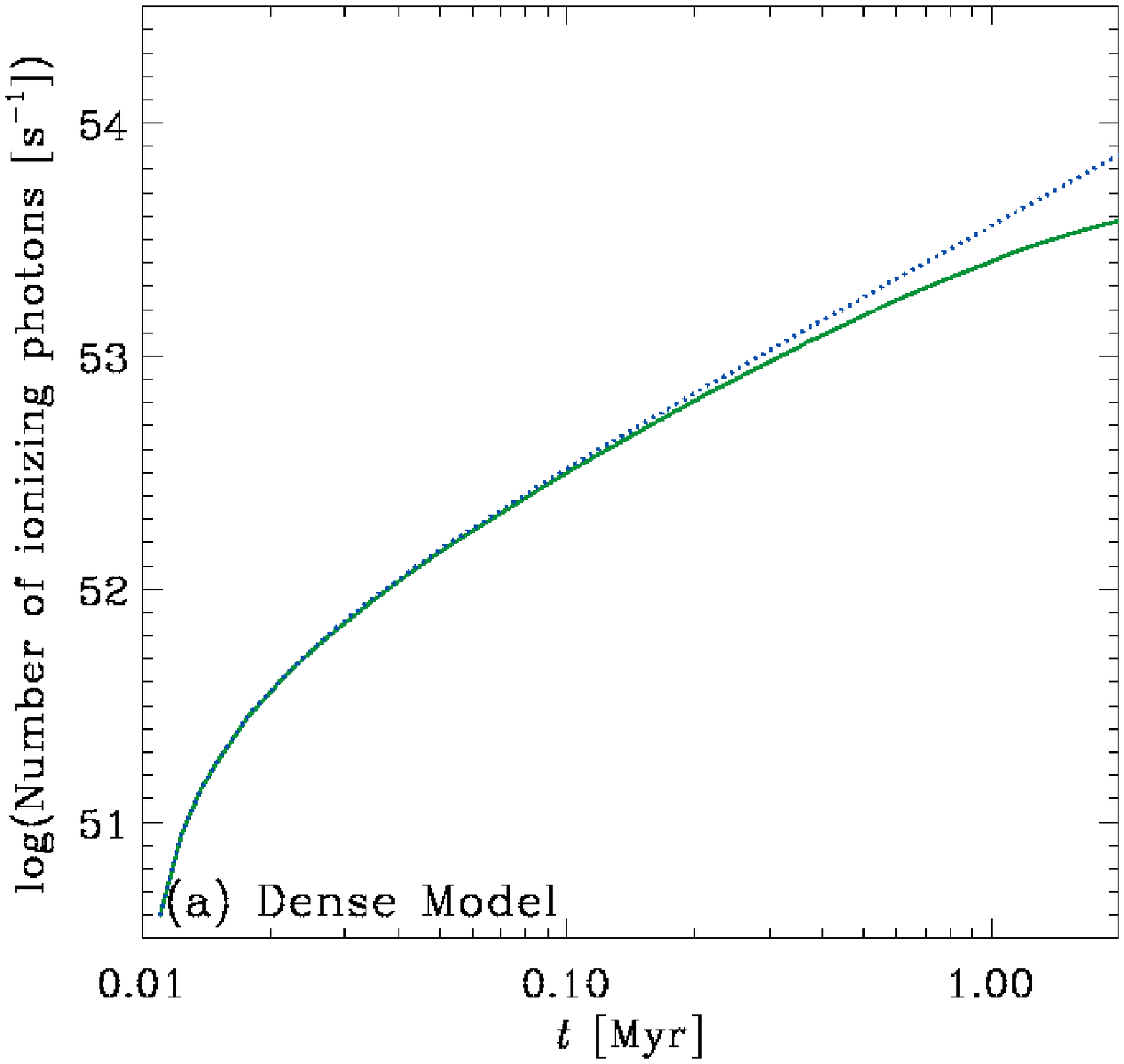}
\hspace{-0.2cm}
\includegraphics[height=4.0cm]{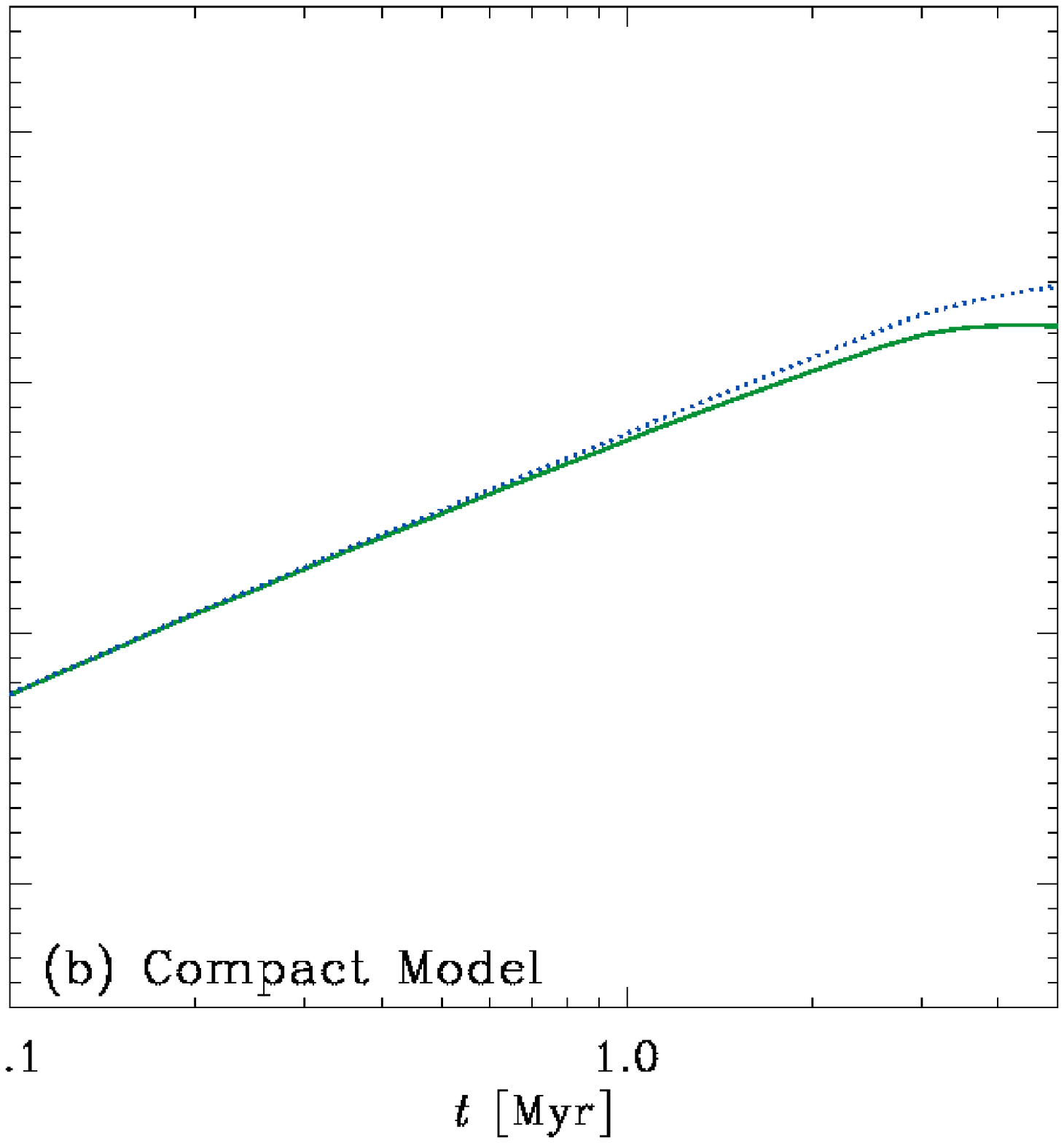}
\hspace{-0.2cm}
\includegraphics[height=4.0cm]{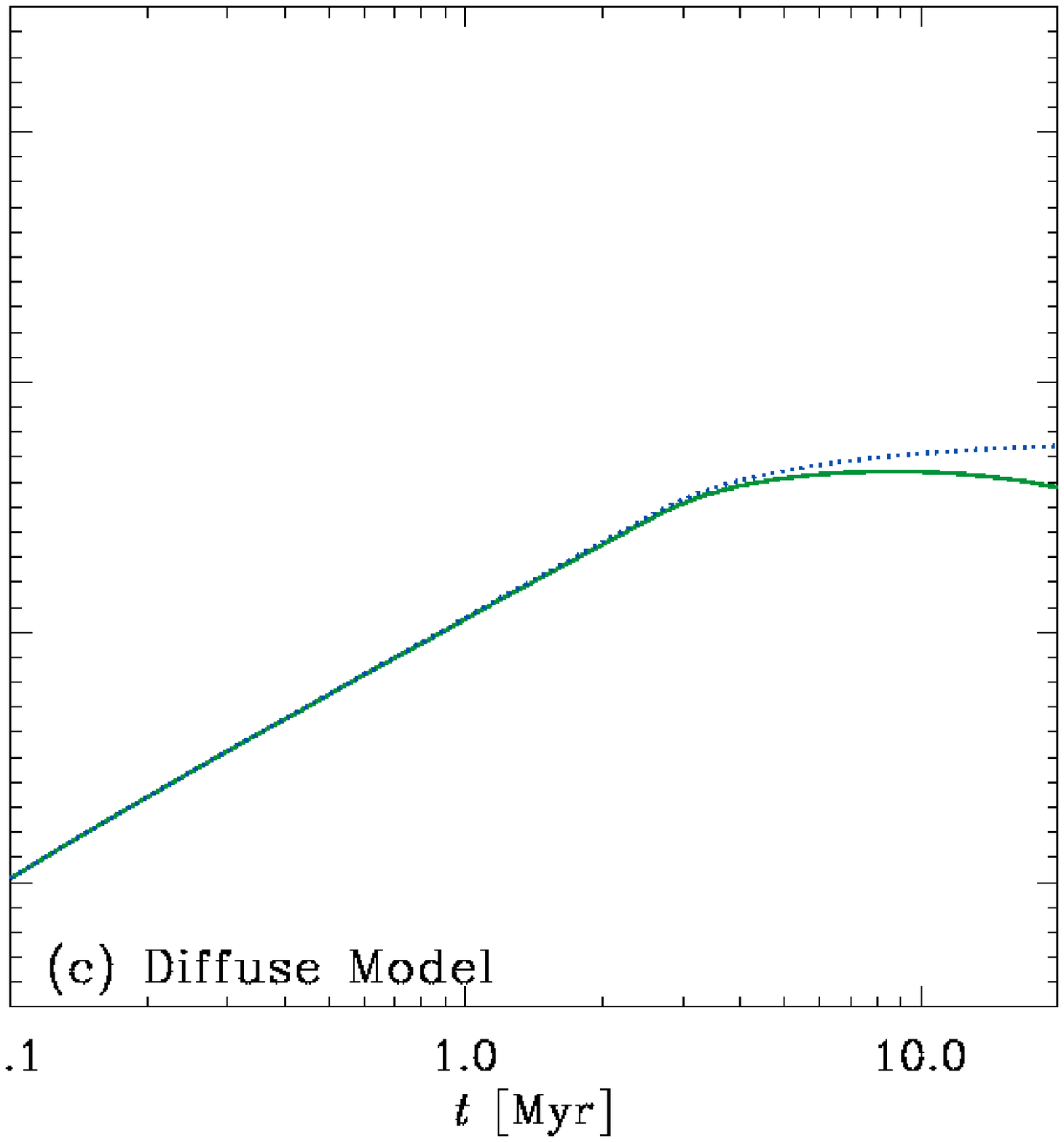}
}
\caption{Time evolution of the number of
ionizing photons emitted per unit time
(\nion). The dense
(Panel a), compact (Panel b), and diffuse (Panel c)
models are
adopted. The solid and dotted lines in each panel
correspond to the exponentially decaying and
constant star formation rates, respectively.}
\label{fig:nion}
\end{figure*}

\subsection{Active and passive star formation}

We have argued in previous papers that star formation at
low metallicities can proceed in two ways:
one ``active'' mode in which stars form in dense, compact
complexes,
and the other ``passive'' mode where star formation occurs
over more diffuse and extended regions.
The size--density relation presented here for extragalactic
\hii\ regions naturally lends itself to the active/passive
dichotomy (see also Hunt et al.\ \cite{hunt-cozumel}).
Dense regions tend also to be compact, while less dense ones
are more extended.
Here we examine these ``active'' and ``passive'' cases.
In HH06, \sbs\ was used as a prototype of the ``active'' mode,
while \izw\ the ``passive'' one.
This representation was based mainly on the radio continuum results
for both galaxies (Hunt et al. \cite{hunt04}; Hunt et al. \cite{hunt05}).
Their linear emission measures differ by 3 orders of magnitude,
and the resulting densities by a factor of 10. 
However, even in the optical, the electron densities inferred
from the [S\,{\sc ii}] line ratios differ by a factor of 5 or so
(\sbs\ has $n_{\mathrm e}\sim400$--$600$\cmcubed,
and \izw\  $n_{\mathrm e}\sim100$--$120$\cmcubed, see
Table \ref{tab:optBCD}).
The \nerms\ of \sbs\ calculated in this paper
(Table \ref{tab:optBCD}) is much lower than that adopted
in HH06 ($n_\mathrm{e}\sim 400$--600 cm$^{-3}$;
Izotov et al. \cite{izotov99}).
As mentioned in Sect. \ref{sec:data}
(see also Kennicutt \cite{kennicutt84};
Rozas et al. \cite{rozas98}), because of clumpiness
the densities from the [S\,{\sc ii}]
lines tend to be significantly higher than the \nerms.
The ``active'' nature of some BCDs therefore may not emerge in the optical,
at least with the rms densities we have adopted here.
However, if a BCD can be classified as ``active'', when
observed in the radio, it would have a compact, dense
(or ultra-dense) \hii\ region (see also Sect. \ref{subsec:density_dep}).

Here, for consistency, we examine ``active'' and ``passive'' cases by
adopting similar initial conditions to those in HH06:
the active case is modelled by the so-called ``compact model'',
and the passive one by the ``diffuse model''.
Appropriate initial densities are
$n_\mathrm{H0}\sim 10^3$--$10^4$ cm$^{-3}$ for
the compact model and $n_\mathrm{H0}\sim 10^2$ cm$^{-3}$
for the diffuse model (HH06). In addition, we also
adopt a denser model as a ``super-active'' class to explain the 
extragalactic radio sample, and call this model the ``dense model''.
For all cases, the gas mass is fixed as $M_\mathrm{gas}=10^7~M_\odot$,
which is similar to the value adopted in HH06.
We calculate the evolution of the size--density relation
of ionized region for the continuous/burst star-formation histories
(Sect.\ \ref{subsec:sfr}).

We also examine the dependence on $\kappa$, by setting 
$\kappa =0$, 0.1, and 1 for dust-free, 
dust-poor, and
dust-rich (Galactic dust-to-gas ratio) cases, respectively.
These cases also correspond to varying degrees of dust
inhomogeneity or gas filling factor, as described in Sect. 
\ref{subsec:kappa}.
The sizes and densities of \hii\ regions 
with a dust content corresponding to $\kappa =0.01$ 
are not significantly affected by dust 
(i.e., the results are similar to the case with $\kappa =0$),
except for the dense model as we discuss later in
Sect.\ \ref{subsec:extinc}.
On the other hand, the effects of dust extinction can be
quite pronounced for $\kappa\gtrsim 0.1$.

\subsubsection{Dense model}

We adopt $n_\mathrm{H0}=10^5$ cm$^{-3}$
($\tff =1.4\times 10^5$ yr) and $M_{\rm gas}=10^7~M_\odot$.
With these assumptions,
the SFR $\psi(t=0)=7.3$\,\msun\,yr$^{-1}$.
First we show the evolution of \nion\ in
Fig.\ \ref{fig:nion}a for
the exponentially decaying and constant SFRs. In both
SFR scenarios, \nion\ increases as time passes and stars form.

\begin{figure*}
\includegraphics[height=3.5cm]{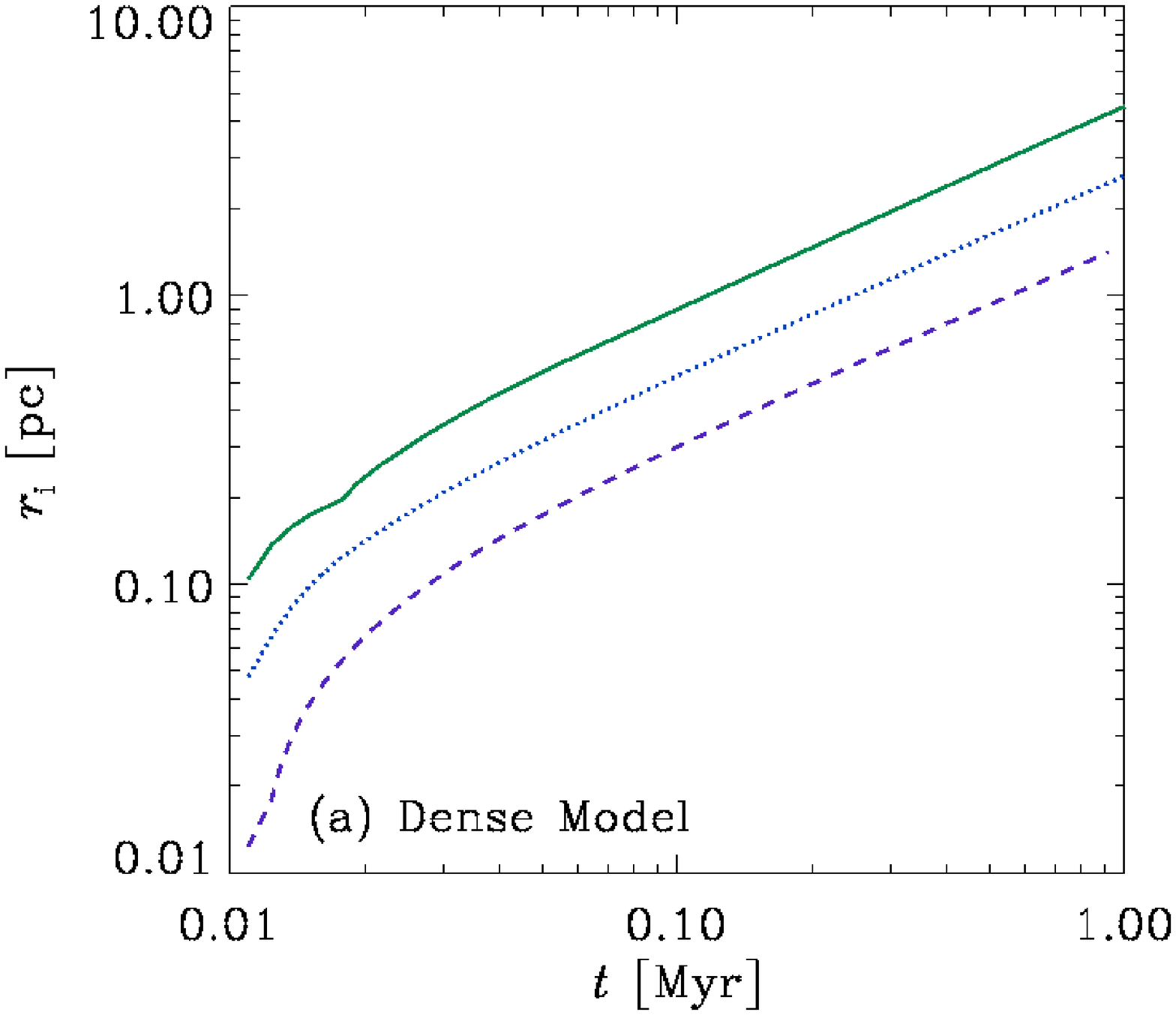}
\includegraphics[height=3.5cm]{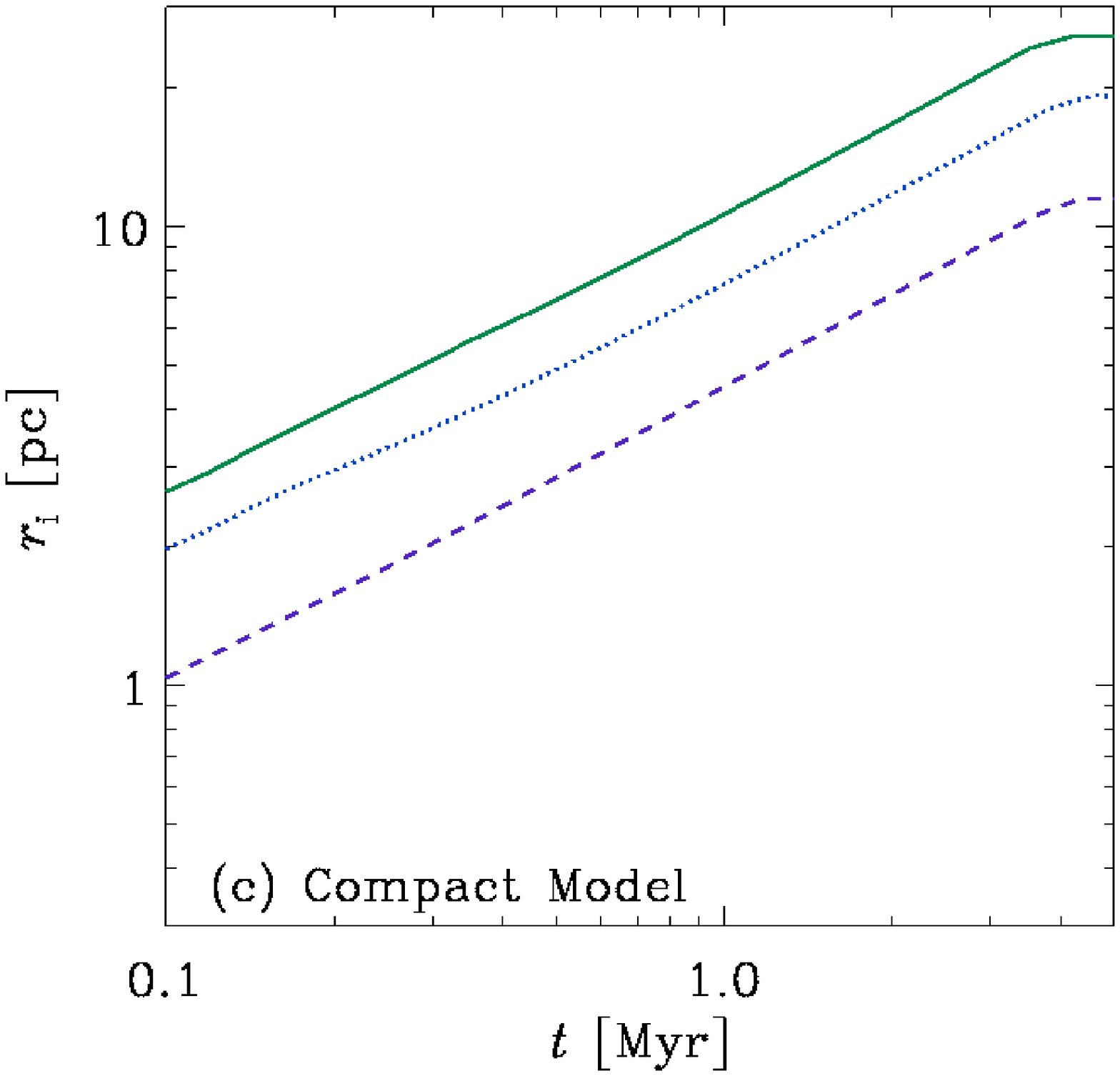}
\includegraphics[height=3.5cm]{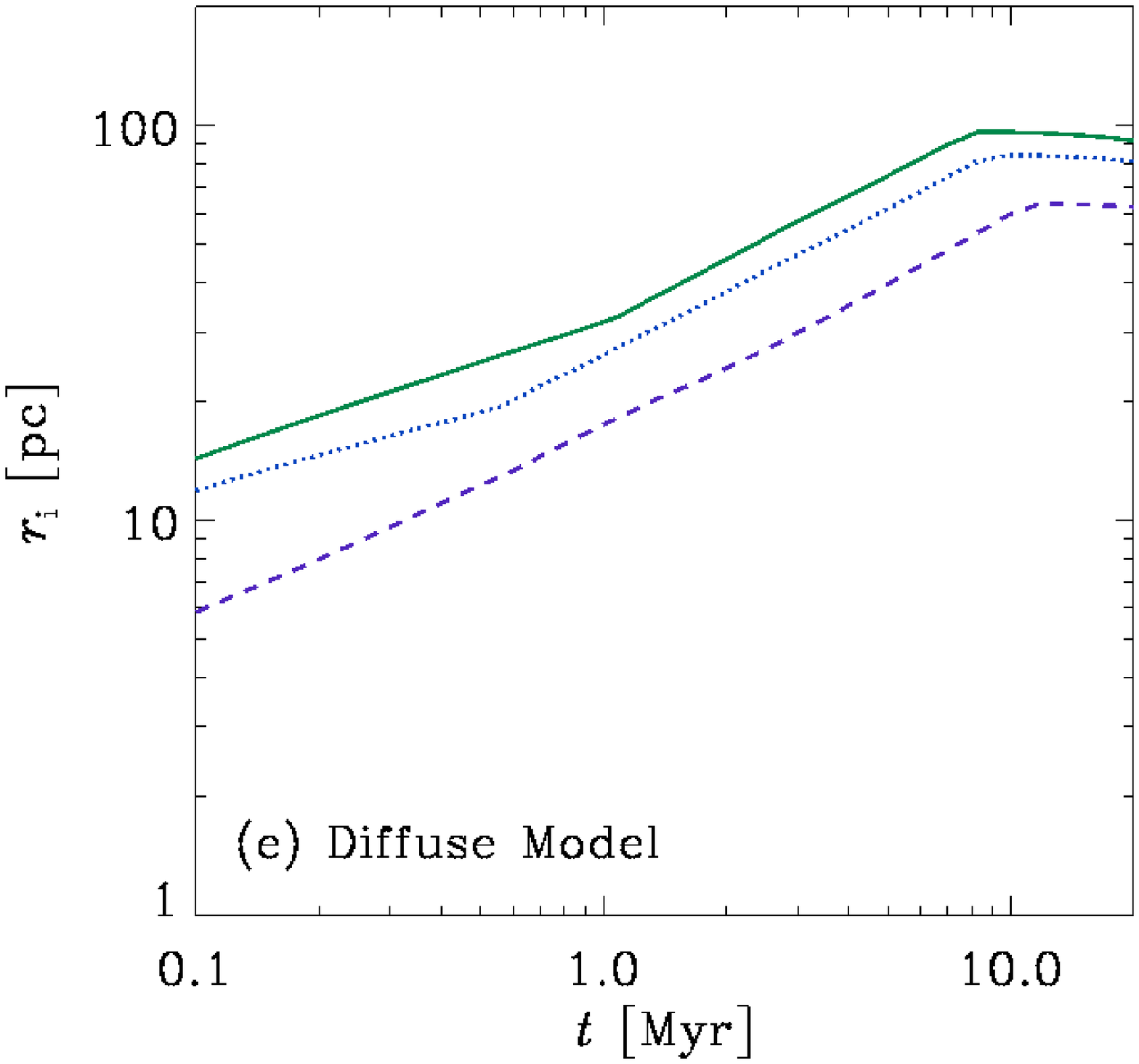}
\\
\sidecaption
\includegraphics[height=3.5cm]{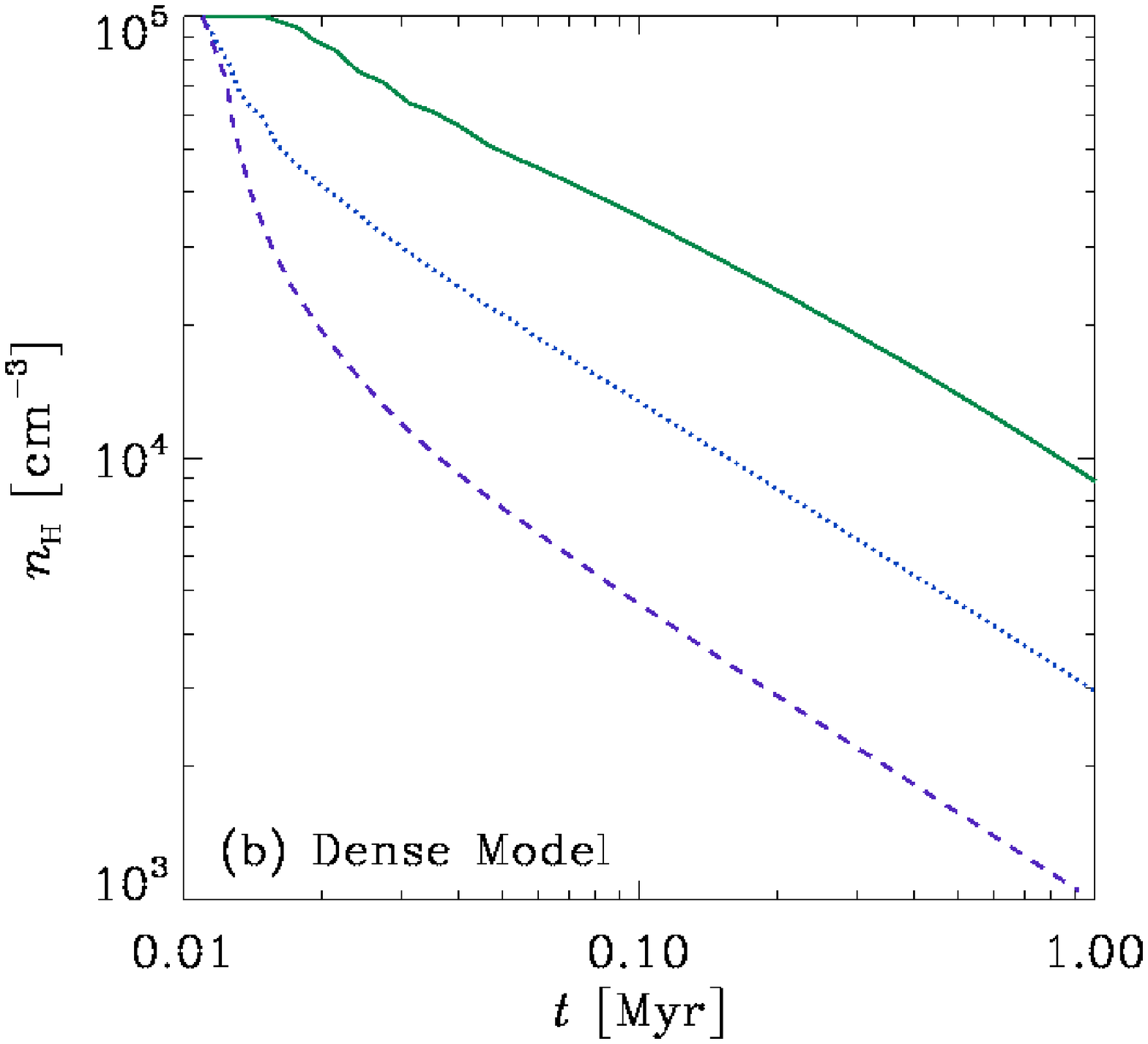}
\includegraphics[height=3.5cm]{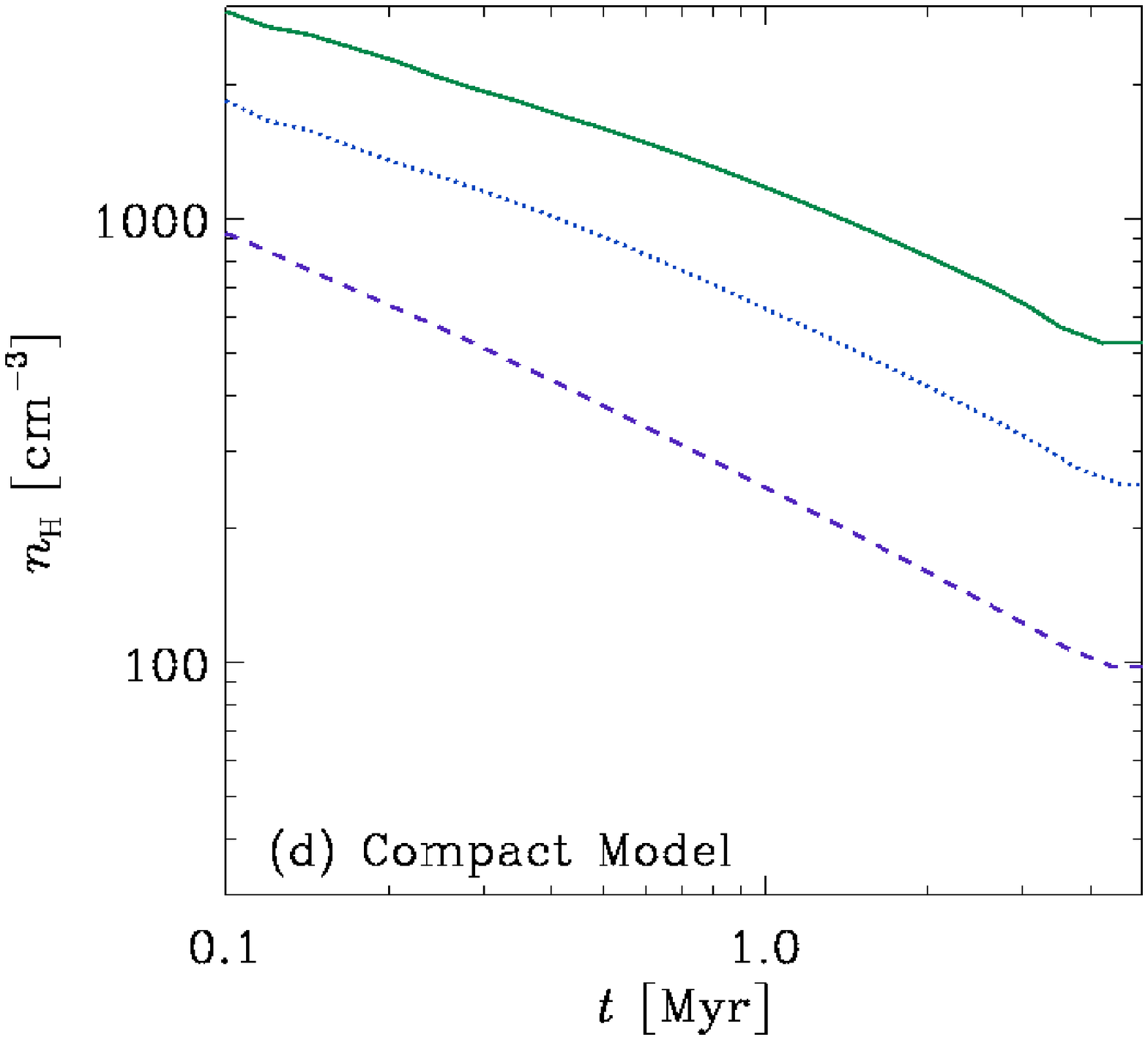}
\includegraphics[height=3.5cm]{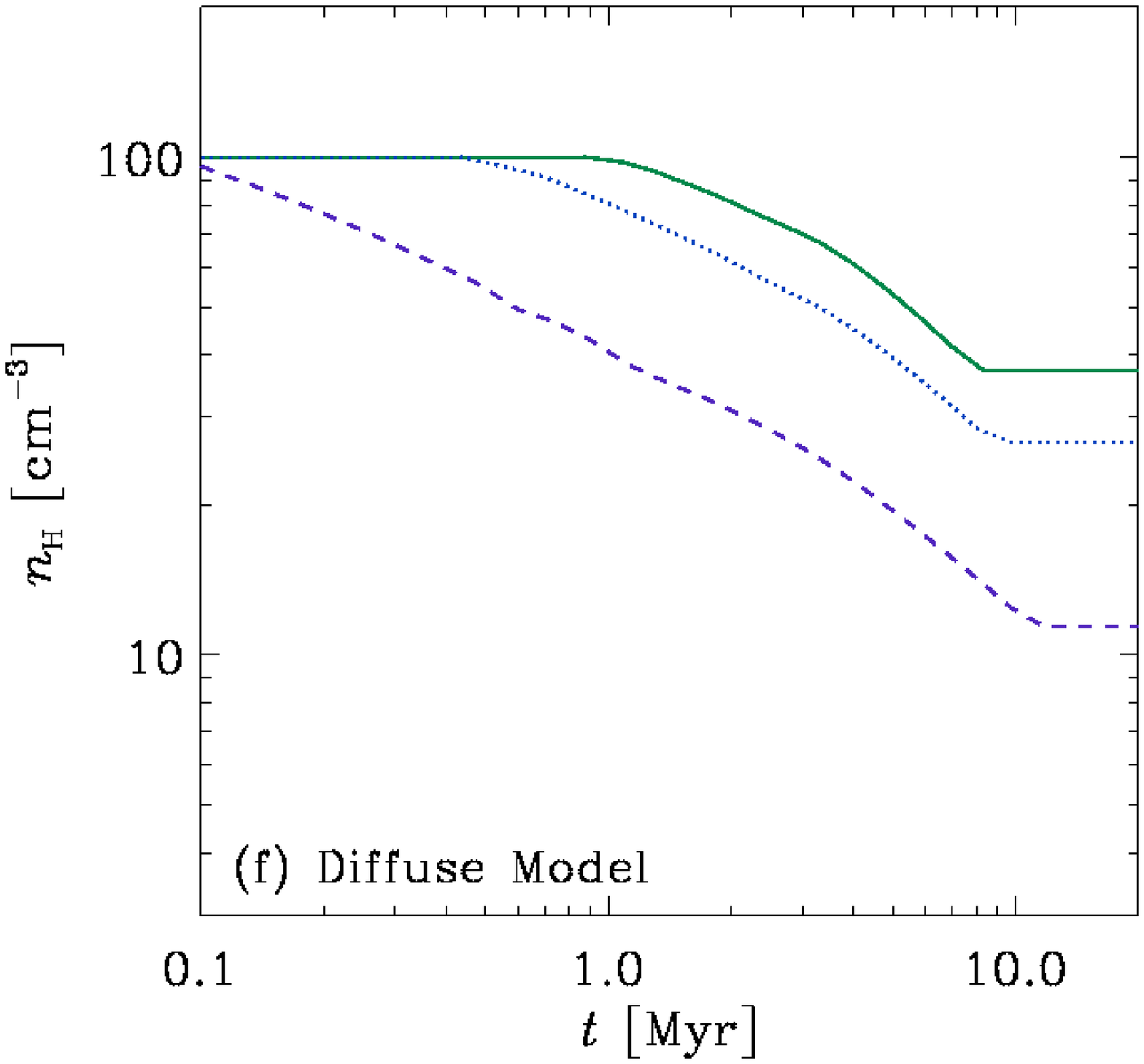}
\caption{Time evolution of the ionization radius
($r_\mathrm{i}$: Panels a, c, and e) and the hydrogen number
density ($n_\mathrm{H}$: Panels b, d, and f). The dense
(Panels a and b), compact (Panels c and d), and diffuse
(Panel c and d) models are
adopted with various dust-to-gas ratio normalized to the
Galactic value, $\kappa$. The solid, dotted, and dashed
lines correspond to $\kappa =0$, 0.1,
and 1, respectively. The exponentially decaying
star formation rate is adopted.}
\label{fig:ri_n}
\end{figure*}

We present the basic evolutionary behavior of the
size and density of \hii\ regions, focusing on
the effect of dust, which is newly incorporated in
this paper.  In Figs.\ \ref{fig:ri_n}a and b, we show
the time variation of the ionization radius and
of the density, respectively, for various $\kappa$
in the exponentially decaying star formation
history. The cases with 
$\kappa =0$ (no dust) are the
same as those investigated in HH06. The general behavior
of pressure-driven expansion is qualitatively similar
in all the dust-to-gas ratios, but quantitatively
different. The ionization radius decreases as
the dust-to-gas ratio increases, because of the
absorption of ionizing photons. The density 
decreases more rapidly in the dustier case,
since the expansion speed normalized to the
ionization radius is larger (for a smaller regions,
the expansion at a given speed decreases the
density more rapidly).  The evolution of $n_\mathrm{H}$
and $r_\mathrm{i}$ is quantitatively similar.

\subsubsection{Compact model}\label{subsec:sbs}

We adopt $n_\mathrm{H0}=3\times 10^3~\mathrm{cm}^{-3}$
($\tff =7.9\times 10^5$ yr) and
$M_\mathrm{gas}=10^7~M_\odot$, as before. 
With these values for the compact model,
the SFR $\psi(t=0)=1.3$\,\msun\,yr$^{-1}$.
First we show
the evolution of \nion\ in Fig.\ \ref{fig:nion}b for
the exponentially decaying and constant SFRs:
Up to $t\sim 1$ Myr, \nion\ increases
as stars form. For the exponentially decaying SFR,
it begins to decrease at $t\sim 5$ Myr because of
the death of massive
stars, while for the constant SFR, it becomes
asymptotically constant because the death of massive
stars is compensated by their continuous birth. 

We present the evolutionary behavior of the
size and density of \hii\ regions in 
Figs.\ \ref{fig:ri_n}c and d
for the exponentially
decaying star formation history.
These models show similar behavior to the dense models.
The increase of $r_\mathrm{r}$
and the decrease of $n_\mathrm{H}$ stop at a later
stage of the evolution because of the exponential
decay of the SFR.

The temporal behavior in the constant star formation is
very similar to  that in the exponentially decaying one.
The difference is that $n_\mathrm{H}$
and $r_\mathrm{i}$ 
continue to decrease and increase, respectively, because the SFR does not 
decay.

\subsubsection{Diffuse model}\label{subsec:izw}

For the initial density and the radius, we assume
$n_\mathrm{H0}=100~\mathrm{cm}^{-3}$
($\tff =4.4\times 10^6$ yr) and
$M_\mathrm{gas}=10^7M_\odot$, as above.
In this case, for the diffuse model,
the SFR $\psi(t=0)=0.2$\,\msun\,yr$^{-1}$.
The evolution of \nion\
is shown in Fig.\ \ref{fig:nion}c for the
exponentially decaying and constant SFRs: around
12 Myr, \nion\ is, respectively,
$6\times 10^{52}$ s$^{-1}$ and
$8\times 10^{52}$ s$^{-1}$. The behavior
of \nion\ in the diffuse models as a
function of time is
qualitatively the same as that in the compact models
(Sect.\ \ref{subsec:sbs}), but with a different SFR.
We present the basic evolutionary behavior of the
size and density of \hii\ regions for the diffuse
models with various dust-to-gas ratios in
Figs.\ \ref{fig:ri_n}c and d for the exponentially
decaying star formation history.
The cases with $\kappa =0$ (no dust) are the
same as those investigated in HH06.
For the dependence on the dust-to-gas ratio,
the same qualitative discussion as in
Sect.\ \ref{subsec:sbs} holds.

The increase of $r_\mathrm{r}$
and the decrease of $n_\mathrm{H}$ stop
for the same reason as in the compact models.
The time evolution in the constant star formation history
is very similar to those in the exponentially decaying
one, but $n_\mathrm{H}$ and
$r_\mathrm{i}$ do not stop decreasing and increasing,
respectively.

\subsection{Evolutionary tracks on the size--density relation}
\label{subsec:density_dep}

\begin{figure*}
\sidecaption
\hbox{
\includegraphics[height=6cm]{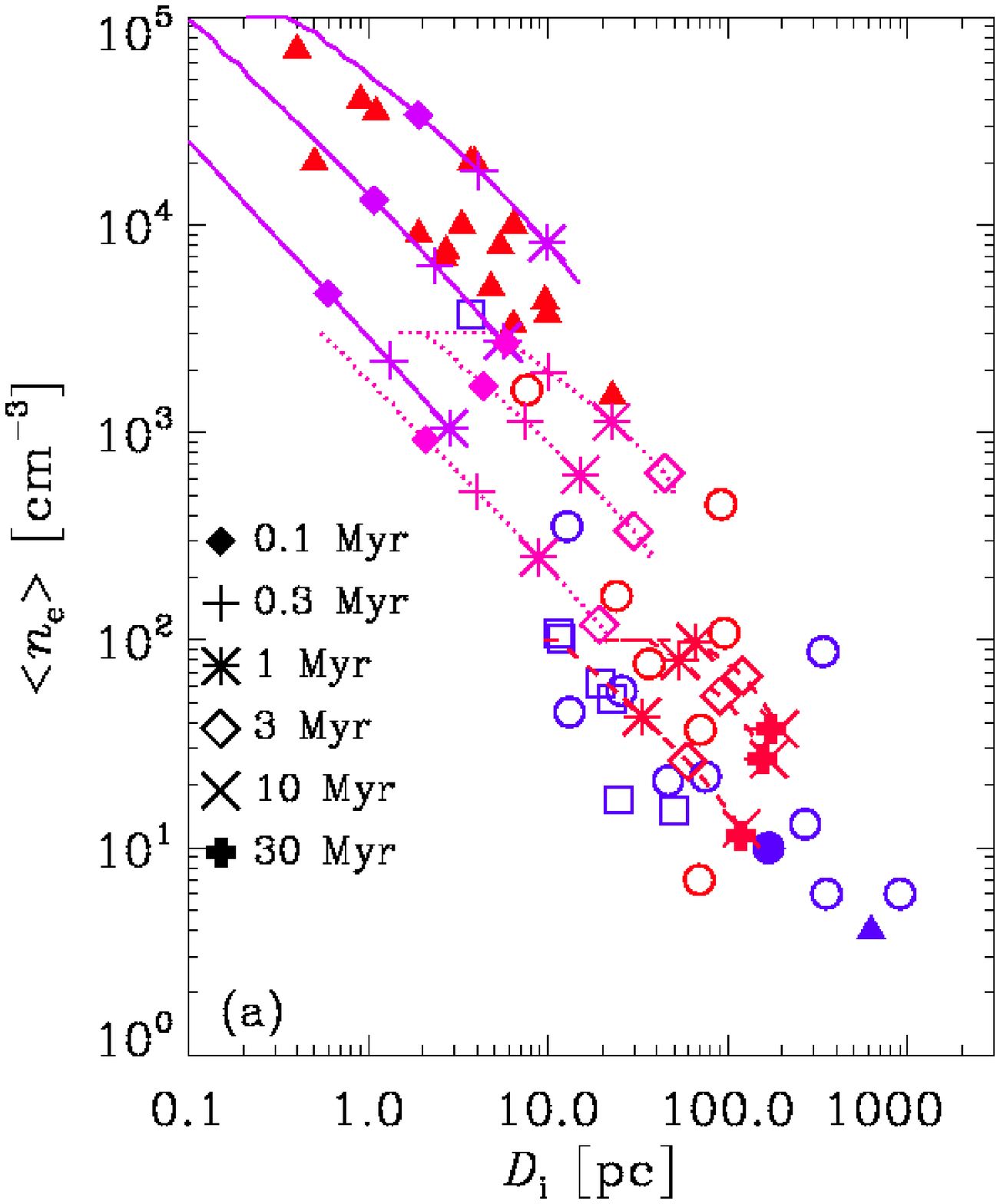}
%\hspace{-0.15cm}
\hspace{-0.2cm}
\includegraphics[height=6cm]{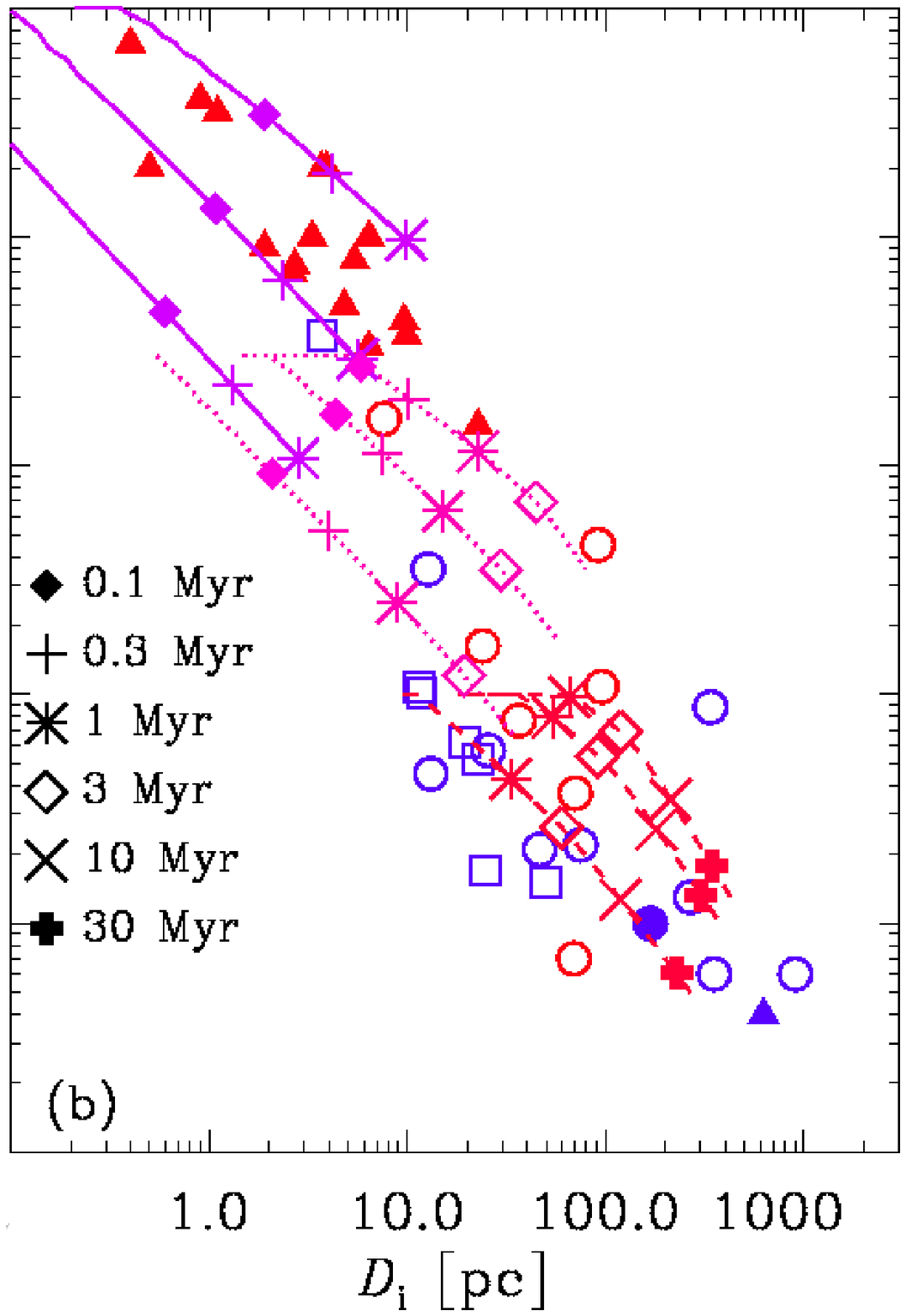}
}
\caption{Relation between the rms electron number density
$\langle n_\mathrm{e}\rangle$ and the ionization diameter
$D_\mathrm{i}$ for {\bf a)} the exponentially decaying
SFR and {\bf b)}
the constant SFR. The dense,
compact, and diffuse models are shown by 
solid, dotted, and dashed lines, respectively,
and the three classes for each model correspond
to $\kappa$\,=\,0.01 (top), 0.1, and 1 (bottom). 
The ages along the model curves are denoted by 
different symbols as marked. 
The data of the radio sample
and the \textit{HST} sample are shown with the
same symbols as in Fig.\ \ref{fig:overall}}.
\label{fig:dwarf_dusty_ev}
\end{figure*}

In Figs.\ \ref{fig:dwarf_dusty_ev}a and b, we show the time
evolution of the ionized region on the size--density
diagram for the exponentially decaying SFR and
the constant SFR, respectively.
For comparison with the data, we show the
diameter of the ionized region, $D_\mathrm{i}$:
\begin{eqnarray}
D_\mathrm{i}=2r_\mathrm{i}\, .
\end{eqnarray}
Below we discuss the results of the three models.

\subsubsection{Specific models}

We observe from Fig.\ \ref{fig:dwarf_dusty_ev}
that the {\it dense models} reproduce the data points
of the {\it radio sample}. This means that the radio sample
can be understood as an extension of the ``active''
mode of star formation toward higher density. However, 
if a Galactic dust-to-gas ratio ($\kappa =1$) is assumed, the
predicted sizes are smaller than those of the radio sample.
This implies that the effects of dust extinction are suppressed
because of an intrinsically low dust content, a relatively high
gas filling factor, or a non-homogeneous dust distribution,
but in a way that is not strictly related to metallicity.

Our models are based on the same gas mass, $10^7$\msun, in the
dense, compact, and diffuse models.
This implies that the star formation
activity traced by the radio sample cannot be
neglected compared with that traced by
the \hst\ (optical) sample.
As shown later in Sect.\ \ref{subsec:extinc},
the fraction of ionizing photons absorbed by
dust tends to be larger in denser regions.
Thus, although they are 
potentially important to the total star formation
activity, 
compact and dense \hii\ regions
such as we have assembled in the radio sample
would tend to be overlooked because of dust extinction.

Fig.\ \ref{fig:dwarf_dusty_ev} shows that
the evolutionary tracks of the compact models
in the size--density diagram explain the upper part
of the \hst\ sample with ages of
$\sim 3$ Myr. Thus, the size and
density of the {\it ``active'' class} can be explained with
the initial conditions of the {\it compact model},
while as shown below,
those of the {\it ``passive'' class} can be reproduced with
the less dense initial conditions of the {\it diffuse model}. 
This is consistent
with the picture that ``active'' and ``passive''
star formation processes originate from dense and diffuse SF
regions, respectively
(Hirashita \& Hunt \cite{hirashita04}; HH06).
%%Our models overpredict the sizes of the ``passive'' BCDs,
%%however, suggesting 
%%that we have overpredicted the typical SFR, either by
%%overestimating the gas mass converted into stars
%%($<10^7~M_\odot$),
%%or overestimating the efficiency ($<$10\%),
%%or both.

\begin{figure*}
\sidecaption
\hbox{
\includegraphics[height=3.925cm]{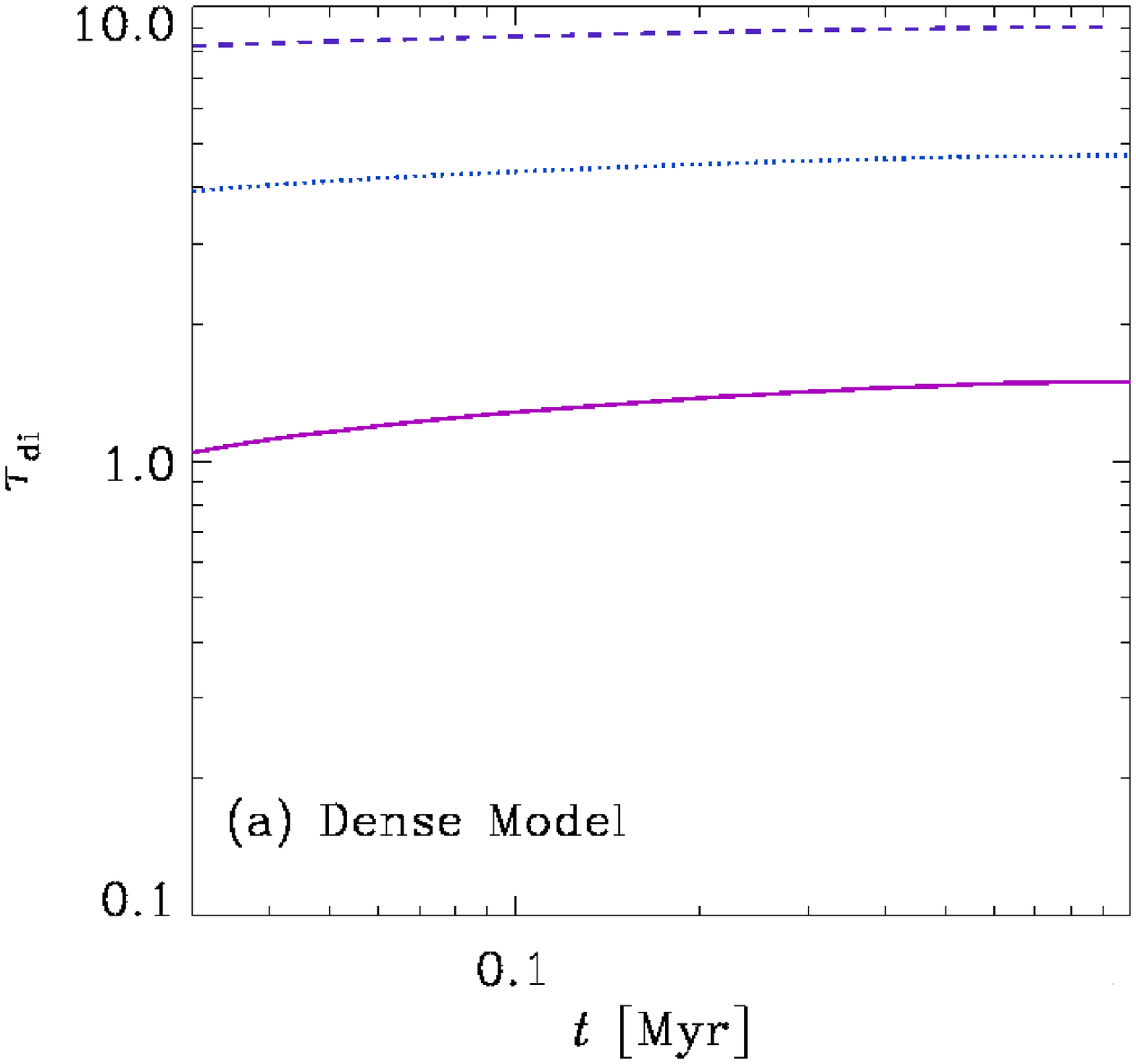}
\hspace{-0.2cm}
\includegraphics[height=3.9cm]{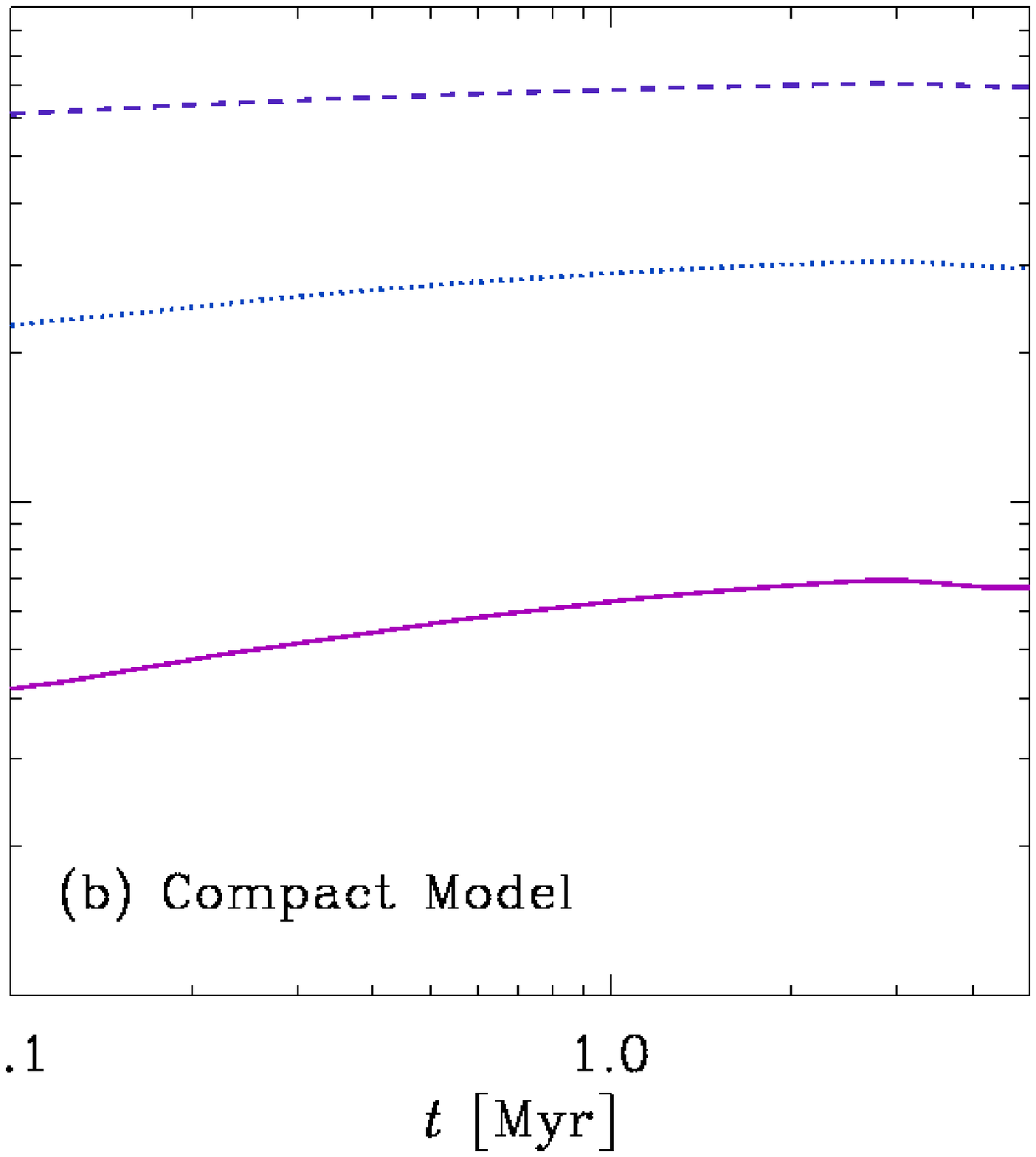}
\hspace{-0.2cm}
\includegraphics[height=3.92cm]{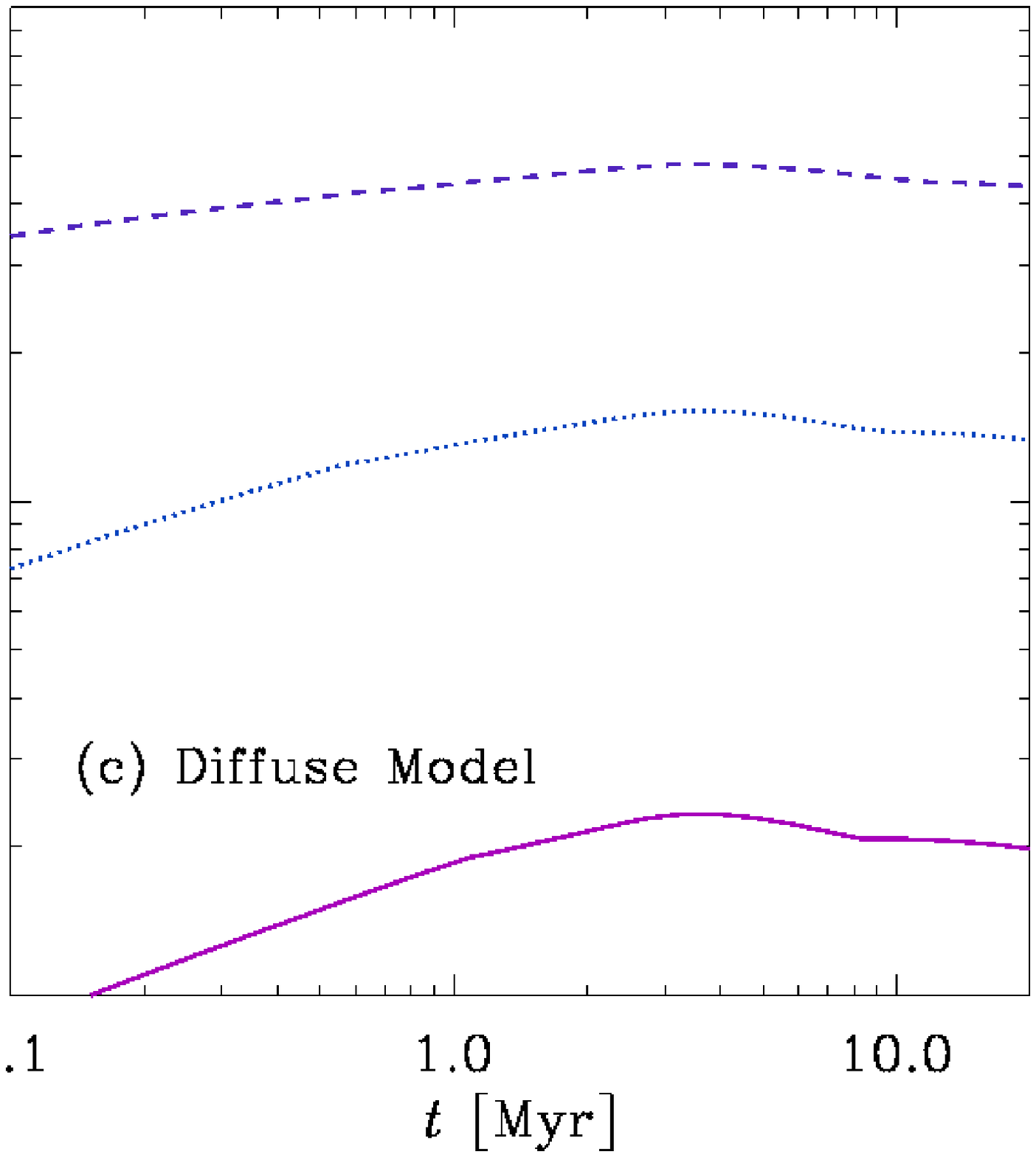}
}
\caption{Evolution of $\taudi$ (the dust optical depth
over the ionization radius). The solid, dotted, and
dashed dotted lines present the results with
$\kappa =0.01$, 0.1, and 1, respectively. The models
adopted are \textbf{a)} dense model,
\textbf{b)} compact model, and
\textbf{c)} diffuse model.
The exponentially decaying SFR is adopted in each
model.}
\label{fig:taudi}
\end{figure*}

The different classes are also distinguished by different
effects of dust extinction.
This is shown in Fig. \ref{fig:taudi}, which presents the
evolution of the dust optical depth, $\taudi$, 
for the different classes of models. 
%We can confirm the trend seen for $f$. 
It can be seen
that $\taudi$ is not a linear function of
$\kappa$; if $\kappa$ increases, $\tau_\mathrm{Sd}$
increases with a fixed values of $n_\mathrm{H}$ and
\nion\ (Eq.\ \ref{eq:tauSd}), but $y_\mathrm{i}$
decreases (Eq.\ \ref{eq:yi_int}). As a result,
the product $\taudi =y_\mathrm{i}\tau_\mathrm{Sd}$
is subject to these competing effects. In particular,
if $\taudi\ga 1$, this nonlinearity becomes significant.
This is why $\taudi$ changes only by a factor of 8
under a change of $\kappa$ by two orders of magnitude in
the dense model. In other words, if dust extinction
dominates, the size--density relation is expected to be
aligned on a strip with $1\la\taudi\la 10$, although
we should also consider the possibility that some
diffuse regions have low dust optical depth if
$\kappa\la 0.01$ is appropriate.

\subsubsection{Summary of the evolutionary tracks}

By examining the above three cases, we have inferred
that very compact dense \hii\ regions such as the radio sample,
compact \hii\ regions in ``active'' BCDs, and diffuse
\hii\ regions in ``passive'' BCDs are different
populations which have emerged because of different initial densities.
%Therefore, in order to trace all the massive stars
%in a BCD, we should consider both compact
%and diffuse ionized regions.
% POTENTIALLY MISLEADING because at the beginning, when we
% define the BCD data, we specifically say that we have
% neglected the diffuse regions, meaning those outside the
% brightest star clusters.
Our models also suggest that denser regions suffer more from the 
effects of dust than less dense ones;
as shown in Fig. \ref{fig:taudi},
``active'' and ``passive'' star formation modes are characterized by 
different dust optical depths.

Contrary to the expectation at
the end of Sect.\ \ref{subsec:static_result},
the radio and \hst\ extragalactic samples cannot be
reproduced with 
a single initial condition. Rather we should consider that
we are seeing an ``envelope'' of individual
evolutionary tracks of \hii\ regions with diverse
initial densities.
The concept of ``envelope'' naturally
suggests that we selectively observe
the end points of the evolution: i.e., we tend to
sample ``active'' SF regions with young ages
of an order of Myr, while we observe ``passive'' regions
with 10 Myr or older ages. In this sense, the observed
size--density relation of extragalactic \hii\ regions
can be taken as an age sequence, although
the relation contains different populations
starting from different initial densities.
The extreme youth of ``super-active'' (extremely dense) regions is 
also supported by the typical rising thermal radio spectrum 
from free-free gas absorption (Cannon \& Skillman \cite{cannon04};
HH06; HH06 also argue that the initial density
is important for the rising spectrum).

As mentioned before (Sect. \ref{sec:scalefree}),
some of the galaxies in the \hst\ sample are also
present in the radio sample, as (ultra-)dense radio \hii\
regions. Hence, it may not only be a question of
evolution, but rather also of what the observations are
probing. It seems that the more finely some \hii\
regions are probed, the smaller and more
dense they appear, even in the same source.
The \hii\ region in NGC\,5253 has a density of 7500--40000\cmcubed\
if observed with a radio interferometer,
but $\sim 300$\cmcubed\ with an optical spectrum. 
The same is true for
He\,2$-$10 (3300--10000\cmcubed\ in the radio and
$\sim 500$\cmcubed\ in the optical;
Vacca \& Conti \cite{vacca92}), 
II\,Zw\,40 (35000\cmcubed\ in the radio and
$\sim 190$\cmcubed\ in the optical), and, as mentioned above,
\sbs\ (1500--7900\cmcubed\ in the radio
and $\sim 500$--600\cmcubed\ in the optical).
Thus, we propose that there is a hierarchy
of size and density. This picture is consistent
with that proposed by Kim \& Koo (\cite{kim01}),
Efremov \& Elmegreen (\cite{efremov98})
and Elmegreen (\cite{elmegreen00}), and will be
further discussed in Sect.\ \ref{subsec:hierarchical}.

\section{Discussion}\label{sec:discussion}

Here we discuss the implications of our results 
for evaluating star-formation activity from ionizing
photons, and examine how the radio and \hst\ samples
fit into the larger picture of star formation locally
and at high redshift.

\subsection{The effects of dust}\label{subsec:extinc}

Even in the dust-poor case with $\kappa\lesssim0.1$,
the absorption of ionizing photons by dust is significant,
%especially in dense regions with $n_\mathrm{H}\ga 10^4$\cmcubed\
especially in dense regions with $n_\mathrm{H}\ga 10^3$\cmcubed\
(see Fig. \ref{fig:dwarf_dusty_ev}).
The effect of dust extinction in the \hst\ sample is
expected to be smaller than in the radio sample,
but it is never negligible according to our results
if $\kappa\ga 0.1$.
Thus, 
dust extinction could constrain the size of \hii\ regions in 
BCDs,
just as it does in Galactic \hii\ regions.

\begin{figure*}
\sidecaption
\hbox{
\includegraphics[height=6cm]{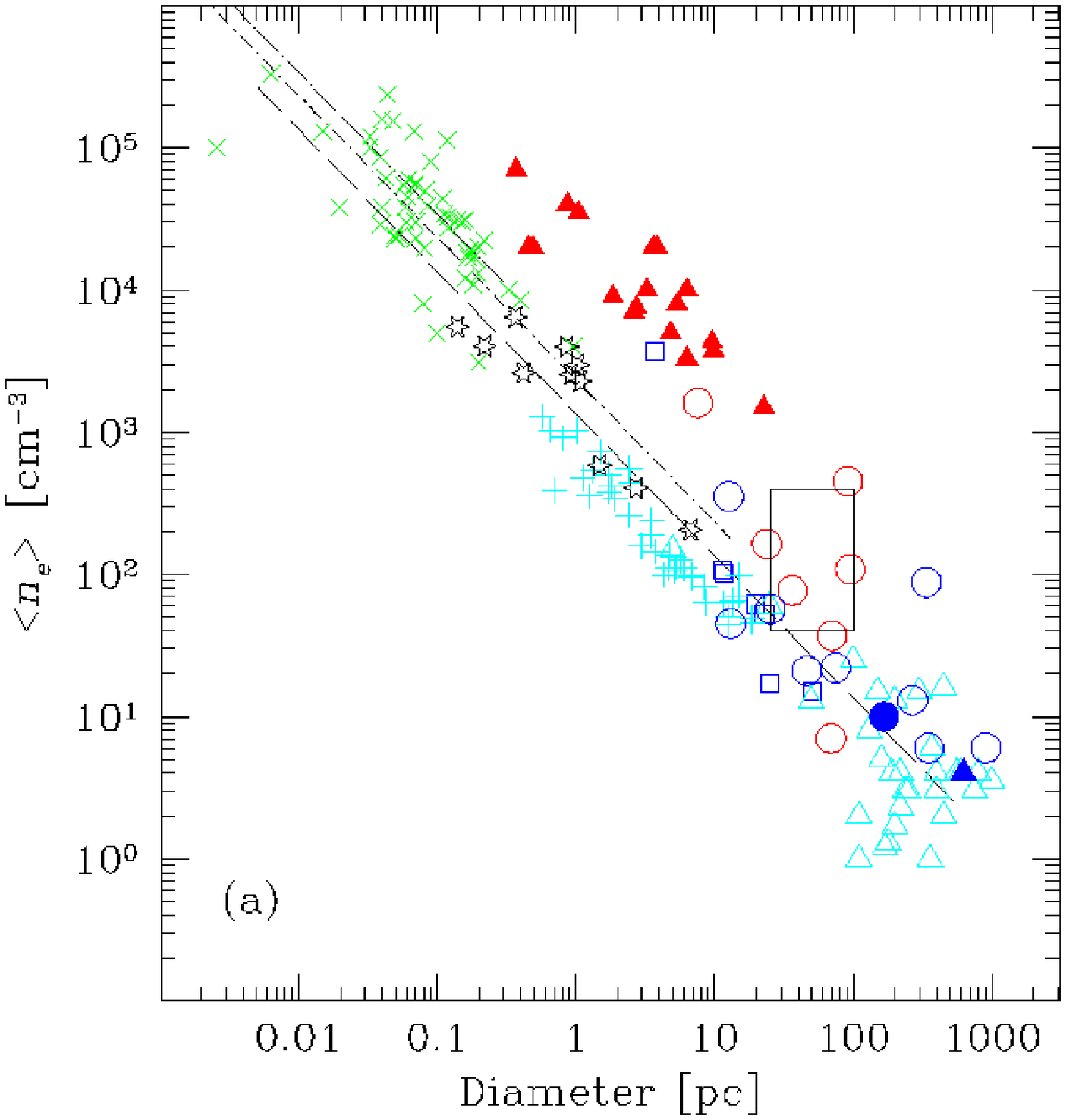}
\hspace{-0.2cm}
\includegraphics[height=6cm]{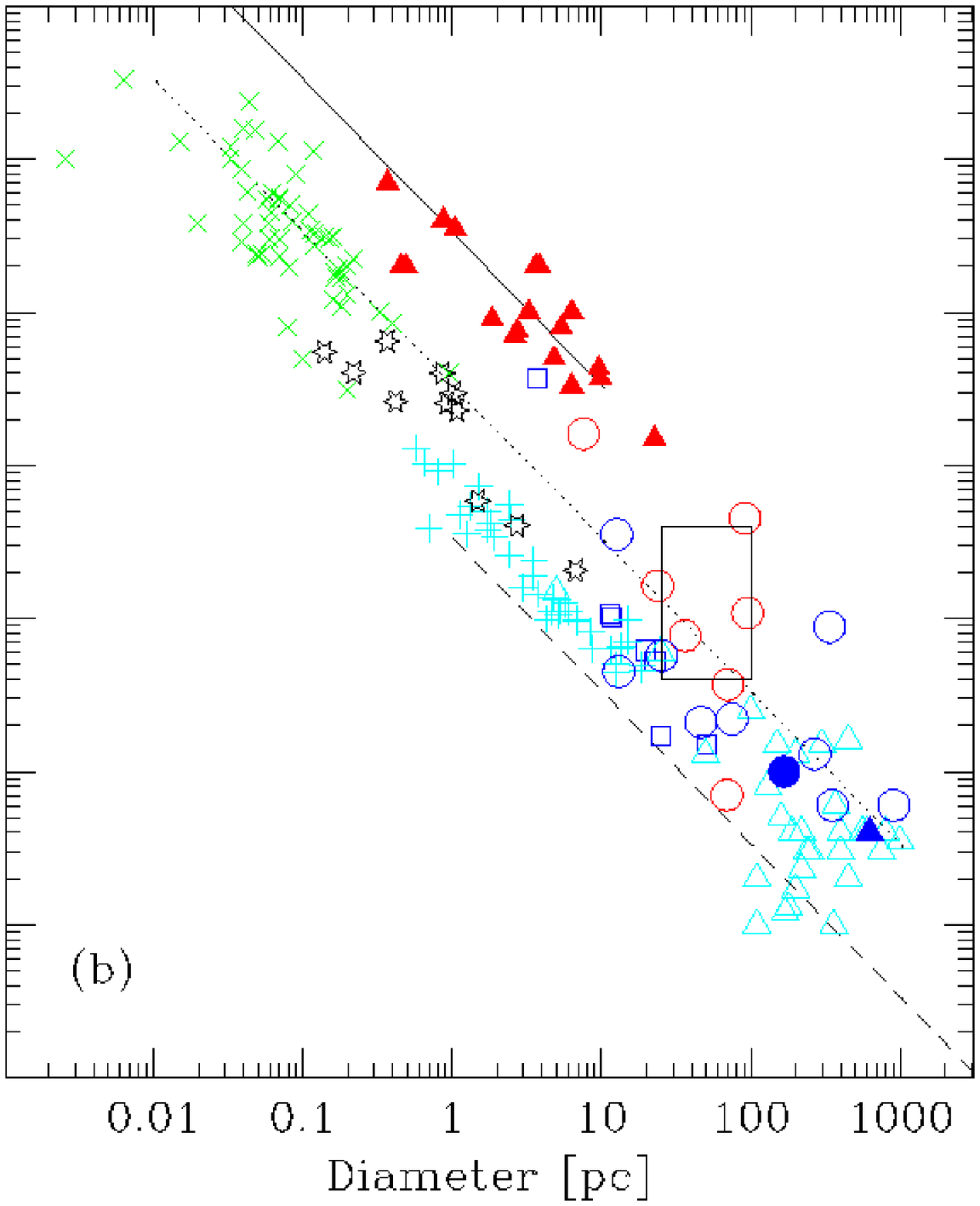}
}
\caption{Size--density relation with a constant dust
optical depth $\taudi$. The data symbols are the
same as in Fig. \ref{fig:overall}. (a) Left panel:
The long-dashed, dot-short-dashed, and dot-long-dashed lines show 
$\taudi =4$, 7, and 10, respectively, with
$\kappa =1$, as predicted by the asymptotic behavior of our models.
(b) Right panel: The solid, dotted, and dashed lines
correspond to $\taudi =1$ with $\kappa =0.01$, 0.1, and 1,
respectively.
\label{fig:dusty_crit}
}
\end{figure*}

The most straightforward way to confirm if dust extinction
significantly alters the size--density relation is to overplot
on the data the dust optical depth calculated by the
models in Sect.\ \ref{sec:static}. For this purpose,
we fix the expression for $\taudi$ (Eq.\ \ref{eq:taudi})
to a constant, and then use the appropriate model 
$n_{\mathrm e}$ and radius (\ri) which would give that $\taudi$
as an asymptotic value.
This is done in Fig.\ \ref{fig:dusty_crit} for
$\taudi =4$, 7, and 10 for $\kappa =1$ and $\taudi =1$ for
for $\kappa =0.01$, 0.1, and 1.
This is a similar diagram to Fig.\ \ref{fig:coldens}, but with the
difference that there we corrected the ionized gas column 
densities assuming a linear dependence on metallicity.
As for Fig.\ \ref{fig:coldens}, $\taudi$ was calculated for densities 
which varied from $10^{-2}$
to $10^6$\cmcubed, and \nion\ from $10^{48}$ to $10^{53}$\,s$^{-1}$;
this is the reason that the lines of constant $\taudi$ do not extend
over all radii and densities.

With a fixed value of $\kappa$, large $n_\mathrm{e}$ and/or large \ri\
are necessary to increase the dust optical depth $\taudi$.
Thus, increasing $\taudi$ moves the line of constant dust optical depth 
upward and to the right. 
However, 
a small value of $\kappa$ with a fixed $\taudi$
shifts the line of constant $\taudi$ in the same direction, 
because for a small value
of $\kappa$, $n_\mathrm{e}$ and/or \ri\ need to increase in order to
keep $\taudi$ constant. Therefore, the line of constant dust optical
depth depends strongly on the adopted value of $\kappa$.

Several salient points emerge from Fig. \ref{fig:dusty_crit}, and from
the comparison of Figs. \ref{fig:dusty_crit} and \ref{fig:coldens}.
First, dust is an important factor in most of the \hii\
regions in our sample, both Galactic and extragalactic ones,
since $\taudi\ga 1$ is indicated.
While in Galactic \hii\ regions, this is well established
(e.g., Gail \& Sedlmayr \cite{gail79}; Melnick \cite{melnick79};
Churchwell et al. \cite{churchwell90}),
in low-metallicity BCDs the effects 
of dust have generally been thought to be negligible.

The second point is that in Fig. \ref{fig:dusty_crit} the data 
are fairly well approximated by different values of constant $\taudi$
(except for the extremely compact and dense radio sample).
Fig. \ref{fig:taudi} shows that $\taudi$ predicted by our evolutionary
models varies by no more than a factor of two over the gas
consumption lifetime.
This relative constancy of $\taudi$ predicted by our models
may be one of the main underlying factors in the size--density
relation observed in \hii\ regions.

Thirdly, the correction for metal abundance applied in
Fig. \ref{fig:coldens} does not align the samples (see Sect. \ref{sec:overall}). 
The only difference between Fig. \ref{fig:dusty_crit}
and Fig. \ref{fig:coldens} is the correction for
metallicity assuming that $\kappa\propto$ metallicity.
The implication is that this correction is not valid, 
namely that the dust-to-gas
ratio is {\it not} linear with metallicity (see also
Lisenfeld \& Ferrara \cite{lisenfeld98};
Hirashita et al.\ \cite{hirashita02}).

Finally, the dense (radio) sample can only be approximated by 
small values of $\kappa\sim 0.01$.
Because the mean metal abundance of the radio sample is $\sim 0.22$\zsun,
this low value of $\kappa$ cannot reflect a low dust-to-gas ratio resulting
from a linear variation with metallicity.
Instead, it must point either to regions evacuated of dust through
stellar winds, or to an intrinsic lack of dust perhaps because of
extreme youth.
In addition, a relatively high gas filling factor or
an inhomogeneous dust distribution with a central cavity could
contribute to the small $\kappa$.
It is likely that all of these alternatives are shaping the properties
of these extreme \hii\ regions.

The generally high values of $\taudi$ which are consistent with our
\hii-region samples suggest the possibility that the size of many \hii\
regions is limited by dust absorption of ionizing photons, rather than the
consumption of ionizing photons by neutral hydrogen.
If dust extinction governs the size of the \hii\
region (we call this situation ``extinction-limited''),
we would naturally expect a constant column density
of \hii\ regions, which is equivalent to a
constant dust optical depth under a fixed mass
absorption coefficient of dust grains.
Natta \& Panagia (\cite{natta76}) also derive
a dust optical depth of order unity in the Lyman
continuum from Galactic \hii\ regions.

\begin{figure*}
\sidecaption
\hbox{
\includegraphics[height=3.9cm]{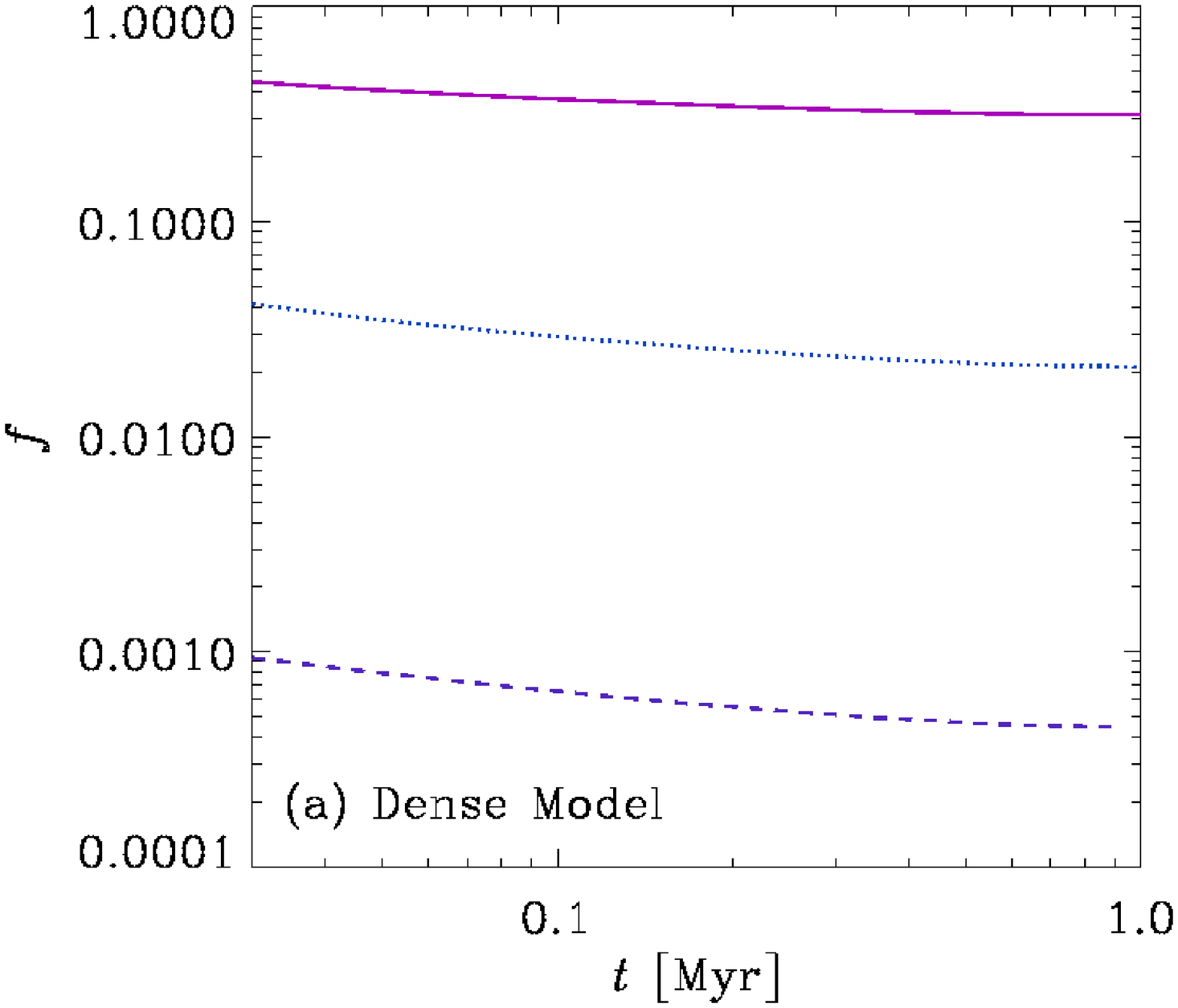}
\includegraphics[height=3.9cm]{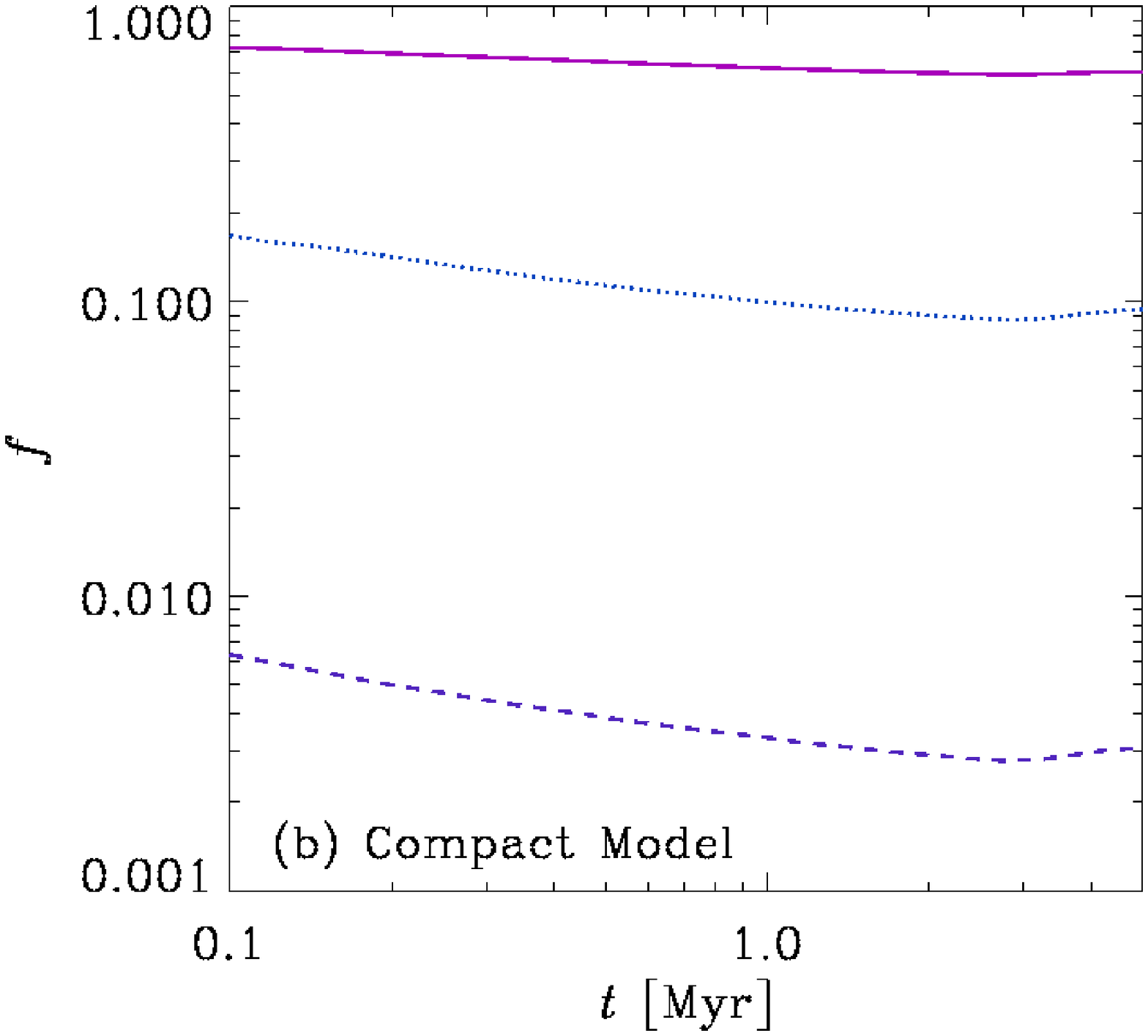}
\includegraphics[height=3.9cm]{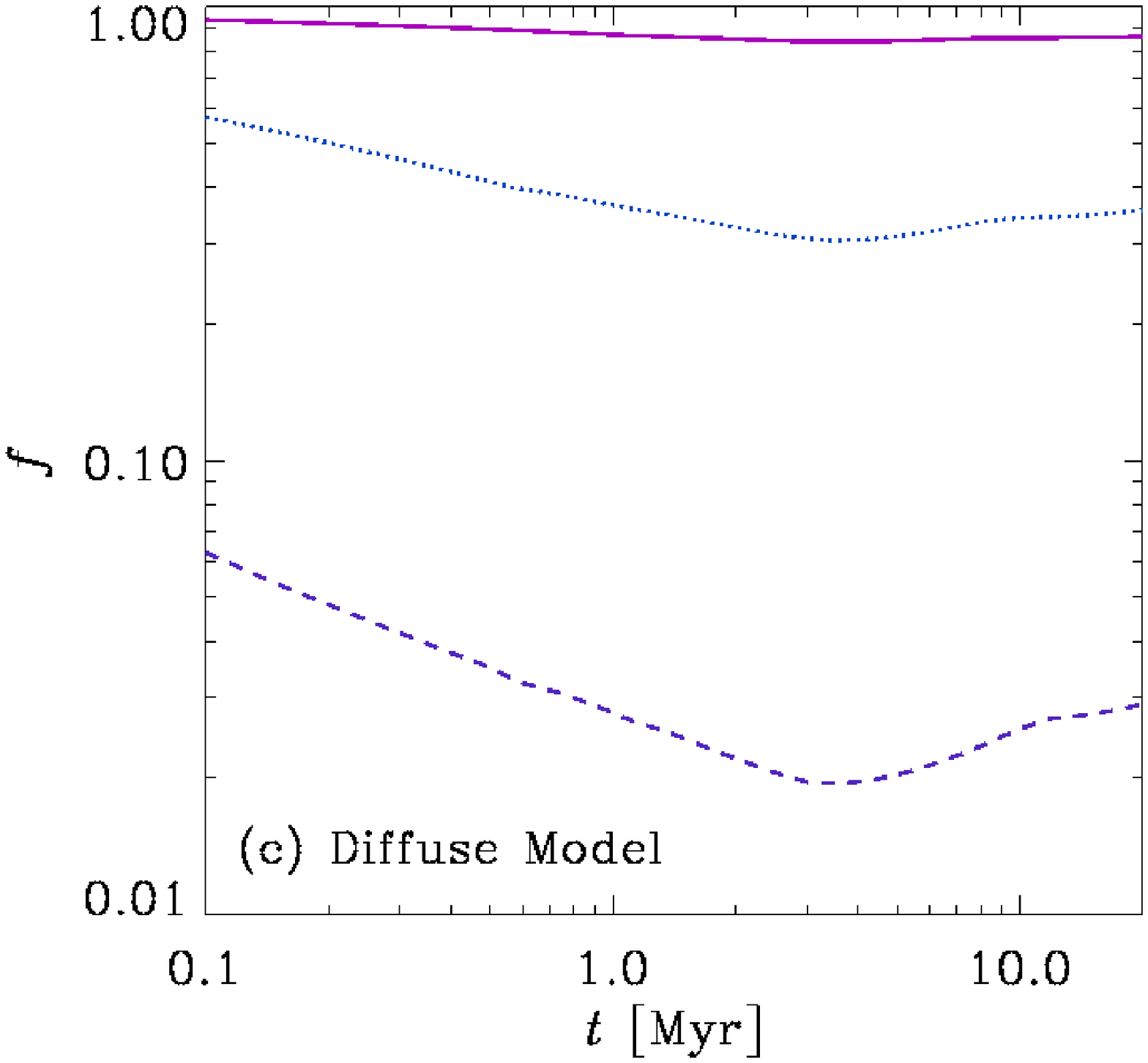}
}
\caption{Evolution of $f$ (the fraction of ionizing
photons absorbed by hydrogen). The solid, dotted, and
dashed dotted lines present the results with
$\kappa =0.01$, 0.1, and 1, respectively. The models
adopted are \textbf{a)} dense model,
\textbf{b)} compact model, and
\textbf{c)} diffuse model.
The exponentially decaying SFR is adopted in each
model.}
\label{fig:f}
\end{figure*}

To further quantify the effects of dust extinction in our
models, we examine the fraction of ionizing photons
absorbed by hydrogen. This fraction is denoted as $f$ and
the fraction of ionizing photons absorbed by dust
becomes $1-f$. By using $y_\mathrm{i}$ determined in
Eq.\ (\ref{eq:yi}), $f$ is expressed as
(Spitzer \cite{spitzer78})
\begin{eqnarray}
f=y_\mathrm{i}^3\, .
\end{eqnarray}
In Fig.\ \ref{fig:f}, we show the evolution of $f$ for
the dense, compact, and diffuse models. It is clear that
more than half of the ionizing photons are absorbed by
dust even with $\kappa =0.01$ in the dense model.
With $\kappa\ga 0.1$, the dust absorption is generally
severe. Thus, we expect that many of
the \hii\ regions in BCDs suffer significant dust absorption of
ionizing photons. This means that
we would grossly underestimate the total mass of massive
stars in compact \hii\ regions in BCDs even from radio
observations. The effect is even more pronounced for hydrogen 
recombination lines.
Such an effect is indeed observed in \sbs\ (Reines et al. \cite{reines08}),
and more work is needed to establish this
for statistically significant samples.

However, as noted in Sect.\ \ref{subsec:taud}, 
inhomogeneities in the dust distribution
affect $f$ ($\taudi$), which may be
larger (smaller) if we consider, for example, a central dust cavity 
(Natta \& Panagia \cite{natta76};
Inoue \cite{inoue02}). 
If such a cavity forms
efficiently on a timescale of $\sim 10^5$ yr in
compact \hii\ regions, $f$ may increase and
the above underestimate of the mass of massive stars
could be less severe. For further qualitative discussion,
one should treat a motion equation of dusty gas, which
is left for future work. 

In summary, the above estimates of $\taudi$ and
$f$ imply that the dust grains play a central role in
determining the ionization radius. In particular,
the small values of $f$ in compact ionized regions
indicate that the effect of dust grains significantly
reduce the number of ionizing photons and the ionization
radius. This means that there could be a large fraction of
massive stars which are not traced with radio observations
or with hydrogen recombination lines.

\subsection{Galactic \hii\ regions}

We have shown that the sequence of
extragalactic \hii\ regions 
cannot be understood
as a sequence with a constant \nion\ but that it can
be interpreted as an ``envelope'' of various
evolutionary tracks starting from diverse initial
densities. This may also be true for Galactic \hii\
regions. Some Galactic \hii\ regions are
associated with a single massive star, but some may be
associated with a forming stellar cluster, for which
we can apply the evolution model described in
Sect.~\ref{sec:model}. This may be the reason
why a similar size--density
relation ($\langle n_\mathrm{e}\rangle \propto D_\mathrm{i}^{-1}$) fits
the Galactic \hii\ regions (Fig.\ \ref{fig:overall}).

In any case, 
in agreement with previous work,
our models with constant $\taudi$ suggest that
dust plays an important role in Galactic \hii\ regions.
As shown in Fig. \ref{fig:dusty_crit}, the more compact \hii\
regions are well fit by $\taudi\sim 10$, while the less
compact ones by $\taudi\sim 2$.
Hence, the Galactic \hii\ region sequence could be a sequence
of constant $\taudi$.
Nevertheless, because high values of $\taudi$ are only possible
with high densities and large \nion, it could be that the 
most obscured regions are also the most intrinsically luminous,
as suggested above.

\subsection{``Active'' and ``passive'' BCDs}

In Sect.\ \ref{sec:constraint}, we have shown that it is
not possible to reproduce the entire sample of extragalactic \hii\
regions with a single initial condition. Indeed, 
the two ``active''/``passive'' star-formation modes proposed by
Hunt et al.\ (\cite{hunt-cozumel}) and elaborated upon
in subsequent papers (Hirashita \& Hunt \cite{hirashita04}, HH06),
could naturally result from different initial densities.
%with the conclusion of
%Hirashita \& Hunt (\cite{hirashita04}).
The difference in the gas density (and the compactness)
explains the variation in far-infrared luminosity and
dust extinction
(Takeuchi et al.\ \cite{takeuchi03};
Hirashita \& Hunt \cite{hirashita04};
Takeuchi et al.\ \cite{takeuchi05}) and in the
molecular fraction (Hirashita \& Hunt \cite{hirashita04})
between the two classes.
We had also hypothesized
that the ``active'' class could be associated with
the existence of SSCs. 
This is supported by the
sizes and densities of the known SSCs in the Antennae
galaxies (Gilbert \& Graham \cite{gilbert07}), which
are well approximated by the compact models 
for the active mode of star formation.

In this paper, we have added the radio sample as a dense
extreme. The radio sample could be called
``super-active'' in the sense that their
\hii\ regions are denser and more compact than even the ``active'' BCDs.
Indeed, our models imply that they constitute a
different population from the \hst\ sample, %for the above
%``active'' and ``passive'' categories, 
and that they are never detected in
the optical because of heavy dust extinction,
even at low values of $\kappa$.
Since a large number of
ionizing photons are absorbed by dust grains
(Sect.\ \ref{subsec:extinc}),
even the luminosity of the radio continuum or radio
recombination lines underestimates the
total mass of massive stars. 

The ``super-active'' mode of star formation seen in the
radio sample could also be consistent
with the idea that dense conditions foster the
formation of SSCs (Billett et al. \cite{billett02}).
In fact, ultra-dense \hii\ regions or radio ``supernebulae''
have been proposed as SSC progenitors, observed in
their very young embedded state 
(Kobulnicky \& Johnson \cite{kobulnickyjohnson99},
Johnson et al.\ \cite{johnson03}, Johnson et al.\ \cite{johnson09}).
This would be consistent with our finding that extreme conditions are
necessary to explain the radio sample.
The young age predicted by our models would also be consistent
with the age of $\la$1\,Myr derived from statistical
considerations by Kobulnicky \& Johnson (\cite{kobulnickyjohnson99}).

\subsection{Density, time, and size in hierarchical star formation}
\label{subsec:hierarchical}

\begin{figure}
\includegraphics[bb=18 216 556 648,width=0.5\textwidth]{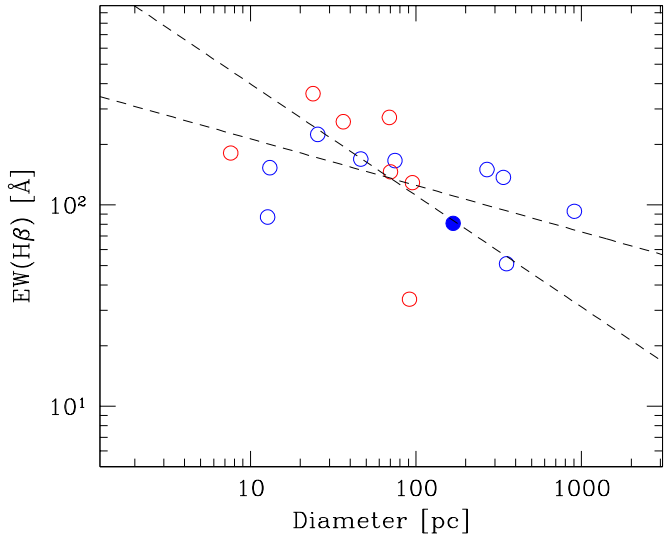}
\caption{H$\beta$ equivalent widths (\AA) vs. Diameter (pc) of the SF
region in the \hst\ sample.
Points are shown as in Fig. \ref{fig:densities}.
The regressions correspond to the best-fit lines with 
slopes of $-0.23$ (minimizing perpendicular distance from the line), 
and $-0.55$ (OLS bisector).
}
\label{fig:ewage}
\end{figure}

Elmegreen and collaborators have proposed that there is 
a timescale--size hierarchy in SF
regions (Efremov \& Elmegreen \cite{efremov98},
Elmegreen \cite{elmegreen00}, and references therein). 
They found a relation between the time difference ($\Delta t$) and 
size and/or separation ($S$) of clouds and star clusters.
We have attempted to assess this trend for the \hst\ sample
in Fig. \ref{fig:ewage}
by plotting the H$\beta$ equivalent width (EW(H$\beta$))
against the diameter of the ELC as described above.
After about 3.1\,Myr, EW(H$\beta$) linearly decreases with starburst
age (Leitherer et al. \cite{sb99}), so we might expect a trend
between age and size to be traced by EW(H$\beta$) and ELC diameter.
The figure shows a weak negative correlation, formally $\gtrsim2\sigma$.
The two dashed lines, correspond to different methods
for determining slopes (Isobe et al. \cite{isobe90}):
minimizing the perpendicular distance to the line gives
$\mathrm{EW(H\beta)}\propto D^{-0.23\pm0.1}$,
and the ordinary-least-squared (OLS) bisector, 
$\mathrm{EW(H\beta)}\propto D^{-0.55\pm0.1}$.
The fitted intercepts at zero size correspond roughly to $\sim 1000$\AA, 
roughly the EW(H$\beta$) ``plateau'' for ages younger than $\sim 3.1$\,Myr
at oxygen abundances 0.05\,\zsun.
The fitted power-law indices for the \hst\ sample
are very close to that found by
Efremov \& Elmegreen (\cite{efremov98}) and Elmegreen (\cite{elmegreen00})
of $\Delta t\propto S^{0.33-0.5}$ for molecular clouds and star clusters in the LMC. 
Hence, even in extragalactic \hii\ regions, there is some (albeit weak) indication that
time and size are related, perhaps because of the imprinting of turbulent molecular cloud
fragmentation on the star clusters responsible for ionizing the \hii\ gas. 

Indeed, the size--density relation of \hii\ regions, 
$n_\mathrm{e}\propto D_\mathrm{i}^{-1}$,
may be compatible with the above time--size relation. 
Considering that the timescale of star formation
scales with the free-fall timescale
(see Sect.\ \ref{subsec:sfr} and Elmegreen \cite{elmegreen00}), 
it may be reasonable to assume a scaling relation
$\Delta t\propto n_\mathrm{e}^{-1/2}$. Combining
these two density scaling relations, we obtain
$\Delta t\propto D_\mathrm{i}^{1/2}$, as long 
the typical size or separation of SF
regions is determined by the size of ionizing
radius, $S\sim D_\mathrm{i}$. We would thus obtain
the same scaling relation as shown by
Efremov \& Elmegreen (\cite{efremov98}) and
Elmegreen (\cite{elmegreen00}). 

Although our data for \hii\ regions cannot be used to
determine star formation activity from
small scales to large (galactic) scales
(e.g., Elmegreen \cite{elmegreen00}), 
we would argue that pressure-driven expansion 
in \hii\ regions and ionization of
the surrounding medium can play a
significant role in determining the typical
extension or separation of SF complexes.
Considering that the dust optical depth 
affects the ionization radius of \hii\ regions,
it is also possible that dust could be a main
factor for determining the typical
extension or separation of SF regions.

\subsection{Implications for high-redshift star formation}

The number density of gas in high-redshift primeval
galaxies is theoretically considered to be
$\ga 10^3$ cm$^{-3}$
(Norman \& Spaans \cite{norman97}). Therefore, the
SF regions in such galaxies
may be expected to mimic our compact models. This means
that a significant fraction of ionizing photons can
be absorbed by dust grains.
According to Fig.\ \ref{fig:f},
roughly 1/3 of the ionizing photons are absorbed
in the compact model even when $\kappa =0.01$.
A pair-instability supernova originating
from a Population III star is expected to supply $\sim 10M_\odot$
of dust grains (Nozawa et al.\ \cite{nozawa03};
Schneider et al.\ \cite{schneider04});
if the baryonic mass of the first object is $\sim 10^5M_\odot$
(Tegmark et al.\ \cite{tegmark97}), the dust-to-gas
ratio becomes $\sim 10^{-4}$ ($\kappa\sim 10^{-2}$).
Thus, $\kappa\sim 10^{-2}$ is reached relatively quickly, and
dust extinction becomes important soon after
the death of the first Population III stars.
Our results also suggest that the radiative transfer
of ionizing photons at high $z$ is strongly affected
by dust extinction. Thus, the cosmic reionization
history should be reconsidered by taking into
account the effects of dust extinction.

It is also interesting to consider the starburst populations
at lower (but still high)
redshifts ($z\la 5$). Although 
resolving individual dense \hii\ regions is impossible in high-$z$
galaxies, 
%%it is worth investigating giant \hii\ regions created by 
an intense starburst at high $z$ would be expected to
produce several giant \hii\ regions or massive ELCs.
For example, let us
assume a star-forming region whose gas mass is three
orders of magnitude larger than that assumed for the
BCDs (i.e., $M_\mathrm{gas}=10^{10}M_\odot$; note that
this is not the total gas mass in the entire galaxy but
the gas mass involved in the current star formation episode).
With $n_\mathrm{H0}\sim 100~\mathrm{cm}^{-3}$
(typical density for the diffuse model)
we can roughly reproduce a typical SFR
$\sim 200M_\odot~\mathrm{yr}^{-1}$ (Eq.\ \ref{eq:sfr})
for luminous infrared galaxies
(Sanders \& Mirabel \cite{sanders96}). If the exponentially
decaying SFR is adopted, this model after $t\sim 30$ Myr
can also be used to
mimic a typical SFR of more mildly star-forming
populations such as Lyman
break galaxies and Ly$\alpha$ emitters
(e.g., Takeuchi \& Ishii \cite{takeuchi04};
Pirzkal et al.\ \cite{pirzkal07}). Moreover, with
$n_\mathrm{H0}\sim 10^4~\mathrm{cm}^{-3}$, the SFR
roughly becomes a typical value for such extreme
starbursts ($\sim 1000M_\odot~\mathrm{yr}^{-1}$) as are
seen in submillimeter galaxies (e.g.,
Hughes et al.\ \cite{hughes98}).

Because the Str\"{o}mgren radius depends only weakly on
\nion\ (Eq.\ \ref{eq:stromgren}), the size of \hii\
regions is only one order of magnitude larger even when
the SFR (\nion\ is proportional to SFR) increases by
three orders of magnitude. 
Severe dust extinction would lessen the increase of \hii\ region 
size. Nevertheless, this implies that high-$z$
starburst population could produce an immense dense \hii\
region whose size ($\ga 100$ pc) is an order of
magnitude larger than
the local (dense) extragalactic \hii\ regions.

In these massive \hii\ regions, if the \hii\ region
size becomes larger than 100\,pc, pressure-driven
expansion ceases to be important. The reason is as follows:
pressure-driven expansion
occurs on a sound-crossing timescale. Since the
sound speed in \hii\ regions is $\sim 10$ km s$^{-1}$,
the sound-crossing time over a 100\,pc region is
10\,Myr, comparable to the typical lifetime of massive
stars. Thus, such a giant \hii\ region cannot be sustained
long enough for the pressure-driven expansion to
modify the ionized-gas density.
Consequently, the density of such a giant \hii\ region
reflects the initial density averaged over the
current \hii\ region size.

\subsection{Maximum-intensity starbursts}

The ``dense'' (``super-active'') and ``active'' star formation
modes we have modelled far exceed the empirical {\it global} star-formation
intensity limit of $\sim$45\,\msun\,yr$^{-1}$\,kpc$^{-2}$
found by Meurer et al. (\cite{meurer97}).
However, recent observations of submillimeter galaxies
and quasar hosts at high redshift have discovered ``hyper-starbursts''
occurring in small regions of $\la$ 1-3\,kpc in diameter
(e.g., Tacconi et al. \cite{tacconi06},
Walter et al. \cite{walter09}).
These kiloparsec-scale starbursts have SFRs per unit area
on the order of 100-1000\,\msun\,yr$^{-1}$\,kpc$^{-2}$,
comparable to those of our models.
Similar high starburst surface densities are found in Arp\,220
but on spatial scales of a few 100\,pc
(Scoville et al. \cite{scoville97}).
With $10^{10}$\,\msun\ of gas, roughly that in Arp\,220,
these are roughly the spatial extents predicted by our evolutionary
models for massive starbursts, as described in the previous section.

If dust is a principle factor in shaping the observable
properties of \hii\ regions and ELCs, 
then we might expect shorter wavelengths, in particular the
ultraviolet, to be unsuitable for sampling SFRs in such objects.
This would mean that starburst intensity limits both locally
and at high redshift would need to be
reassessed at submillimeter or infrared wavelengths where dust 
reprocessing gives a more accurate picture
(see also Gao \cite{gao08}).

\section{Conclusions}\label{sec:conclusion}

We have investigated the size--density relation of
extragalactic \hii\ regions, focusing on those in BCDs.
Motivated by the similarity of size--density relations of
extragalactic \hii\ regions with Galactic ones, we have 
modelled and examined the size--density relation of ionized regions
by considering the effects of dust, star formation history, and
pressure-driven expansion of \hii\ regions.
The results have been compared with several samples spanning 
roughly six orders of magnitude
in size and density. We have shown that the
entire sample set cannot be understood as an evolutionary
sequence with a single initial condition. Rather,
the size--density relation reflects a sequence with
different initial gas densities. Thus, 
a hierarchical structure of SF
regions with various densities is implied.

We have also found that the size of extragalactic
\hii\ regions is ``dust-extinction limited'', in the sense 
that the dust absorption of ionizing photons is significant. 
This naturally explains the observed size--density
relation of \hii\ regions as following a constant column
density of ionized gas, if the dust-to-gas ratio in
\hii\ regions is constant. The
dust extinction of ionizing photons is particularly
severe over
the entire lifetime of the compact radio sample
%with typical densities of $\ga 10^4$\cmcubed. 
with typical densities of $\ga 10^3$\cmcubed. 
This means that the compact radio sample constitutes a
different population from the optical samples and that
star formation activity in such dense regions would be
underestimated or missed entirely if we use the emission from \hii\
regions (hydrogen recombination lines, free-free
continuum) as the sole indicators of star formation rate.

\begin{acknowledgements}

We would like to humbly dedicate this paper to the memory of 
Prof. Edwin Salpeter, who provided precious insight during its 
development.
We are also grateful to B. G. Elmegreen, R. C. Kennicutt,
and A. K. Inoue for stimulating discussions on
the properties of \hii\ regions. 
Finally,
we thank the anonymous referee for useful comments which 
substantially improved the paper.
This research has made use
of the NASA/IPAC Extragalactic Database (NED), which is operated
by the Jet Propulsion
Laboratory, California Institute of Technology, under contract
with the National Aeronautics and Space Administration (NASA).

\end{acknowledgements}

\clearpage

\clearpage

  \begin{table*}
     \caption[]{Radio sample}
        \label{tab:radioBCD}
\begin{tabular}{@{}lcccccc@{}}\hline
Name & $d$ & $n_\mathrm{e}$ & $D$ &
$12+\log\mathrm{(O/H)}$ & ref.$^\mathrm{a,\, b}$ \\
 & (Mpc) & (cm$^{-3}$) & (pc) &
\\ \hline
He\,2$-$10 & 10.4 & 10000 & 6.4 & 8.06 & 1, 9 \\
        &      & 3300  & 6.4 &      & 1 \\
I\,Zw\,18 & 14.6 & 4 & 626 & 7.17 & 3, 10 \\
II\,Zw\,40 & 11.1 & 35000 & 1.1 & 8.13 & 3, 10 \\
M\,33     & 0.84$^\mathrm{c}$ & 3700  & 3.7 & 8.78 & 5, 17 \\
Mrk\,8    & 52.5 & 20000 & 3.7 & 8.51 & 4, 13 \\
Mrk\,33   & 25.5 & 9000  & 1.9 & 8.40 & 4, 14 \\
Mrk\,1089 & 54.6 & 20000 & 3.9 & 8.04 & 4, 12 \\
Mrk\,1236 & 29.6 & 20000 & 0.5 & 8.07 & 4, 16 \\
NGC\,4214n & 3.8 & 70000 & 0.4 & 8.22 & 4, 11 \\
NGC\,5253 & 2.5 & 7500 & 2.7 & 8.17 & 1, 8 \\
         &     & 40000 & 0.9 &   & 2 \\
NGC\,6946 & 5.6  & 3700  & 9.9 & 9.12 & 5, 15 \\
NGC\,253  & 3.4  & 10000 & 3.3   & 8.88 & 5, 17 \\
Pox\,4    & 52.5 & 7000 & 2.6 & 7.95 & 4, 14 \\
% 23/7/2009: now in Local Group sample which is not in a table
% since published numbers
% M\,33$-$N604A     & 0.84$^\mathrm{c}$ & 102 & 11.8 & 8.78 & 6, 18 \\
% M\,33$-$N604B     & 0.84$^\mathrm{c}$ & 52  & 22.8 & 8.78 & 6, 18 \\
% M\,33$-$N604C     & 0.84$^\mathrm{c}$ & 108 & 11.4 & 8.78 & 6, 18 \\
% M\,33$-$N604D     & 0.84$^\mathrm{c}$ & 62  & 19.6 & 8.78 & 6, 18 \\
% M\,33$-$N604E     & 0.84$^\mathrm{c}$ & 17  & 24.8 & 8.78 & 6, 18 \\
% M\,33$-$N604F     & 0.84$^\mathrm{c}$ & 15  & 50.1 & 8.78 & 6, 18 \\
SBS\,0335$-$052   & 53.7    & 1500 & 22.7 & 7.30 & 6, 10\\
SBS\,0335$-$052$-$1   & 53.7    & 4300 & 9.6 & 7.30 & 7, 10\\
SBS\,0335$-$052$-$2   & 53.7    & 7900 & 5.4  & 7.30 & 7, 10\\
Tol\,35   & 25.2 & 20000 & 0.5 & 8.11 & 4, 15 \\
VII\,Zw\,19 & 67.4 & 5000 & 4.8 & 8.71 & 4, 14 \\
\hline
\end{tabular}
\begin{list}{}{}
\item[$^{\mathrm{a}}$] References for the
radio interferometric observations:
1) Mohan et al.\ (\cite{mohan01});
2) Turner et al.\ (\cite{turner00});
3) Beck et al.\ (\cite{beck02});
4) Beck et al.\ (\cite{beck00});
5) Johnson et al.\ (\cite{johnson01});
%6) Churchwell \& Goss (\cite{churchwell99});
6) Hunt et al. (\cite{hunt04});
7) Johnson et al. (\cite{johnson09}).
\item[$^\mathrm{b}$] References for the oxygen abundance:
8) Walsh \& Roy (\cite{walsh89});
9) Vacca \& Conti (\cite{vacca92});
10) Thuan \& Izotov (\cite{thuan05});
11) Kobulnicky \& Skillman (\cite{kobulnicky96});
12) Guseva et al.\ (\cite{guseva00});
13) Shi et al.\ (\cite{shi05});
14) Kunth \& Joubert (\cite{kunth85});
15) Kobulnicky et al.\ (\cite{kobulnicky99});
16) Buckalew et al.\ (\cite{buckalew05});
17) Garnett (\cite{garnett02}).
\item[$^{\mathrm{c}}$] The distance of M33 is
taken from Freedman et al. (\cite{freedman91}).
\end{list}
  \end{table*}

  \begin{table*}
     \caption[]{\hst\ sample }
        \label{tab:optBCD}
{\scriptsize
\begin{tabular}{@{}lcccrrcccrrc@{}}\hline
Name & Dist.\,$^\mathrm{a}$ 
& $12+$ 
& $T_\mathrm{e}$\,$^\mathrm{c}$ 
& \nesii\,$^\mathrm{d}$ 
& \nerms\,$^\mathrm{e}$ 
& \multicolumn{1}{c}{$F_{H\beta}$\,$^\mathrm{f}$} 
& \multicolumn{1}{c}{EW(H$\beta$)\,$^\mathrm{g}$} 
& Filter
&  $D$\,$^\mathrm{h}$ 
&  $D$\,$^\mathrm{i}$ 
& Ref.$^\mathrm{j}$ \\
& (Mpc) & $\log (\mathrm{O/H})\,^\mathrm{b}$ & ($\times10^4$ K) 
& (cm$^{-3}$) 
& (cm$^{-3}$) 
& \multicolumn{1}{c}{($\times$10$^{-14}$}
& \multicolumn{1}{c}{(\AA)} 
&& (pixel) & (pc) \\
&&&&&& erg\,cm$^{-2}$\,s$^{-1}$) \\
(1) & (2) & (3) & (4) & (5) & (6) & (7) & (8) & (9) & (10) & (11) & (12) \\
\hline
\\
Haro\,3        &  18.5 &   8.34 & 1.019 &   180 &    37 &  26.4 &   146 & F606W &   11.6 &   70.2  & 1, 2 \\ 
He\,2-10-A     &  10.4 &   8.34 & 1.120 &   383 &   450 & 261.0 &    34 & F814W &   12.1 &   91.4  & 3, 4, 5\\ 
HS\,0822$+$3542 & 12.6 &   7.46 & 2.000 &   390 &     7 &   0.5 &   272 & F814W &   30.2 &   69.1  & 2, 6, 7, 8 \\ 
II\,Zw\,40       &  11.1 &   8.13 & 1.302 &   190 &    77 &   8.5 &   259 & F814W &   18.0 &   36.4  & 2, 9 \\ 
I\,Zw\,18-NW     & 14.6 &   7.17 & 1.997 &   120 &    10 &   3.2 &    81 & F814W &   35.1 &  167.8  & 2, 10 \\ 
Mrk\,59-1    &  13.4 &   8.01 & 1.354 &    79 &    13 &  15.1 &   150 & F606W &   27.6 &  268.7  & 2, 11\\ 
Mrk\,71-A    &    4.4 &   7.89 & 1.568 &   150 &   163 &  87.3 &   357 & F814W &    7.5 &   23.9  & 10, 12\\ 
Mrk\,209      &   5.2 &   7.81 & 1.561 &   100 &    57 &  14.6 &   224 & F814W &   13.4 &   25.5   & 2, 10 \\ 
Mrk\,930      &   74.6 &   8.06 & 1.230 &    58 &     6 &   6.9 &    93 & F606W &   16.7 &  905.9   & 13 \\ 
Mrk\,996      & 21.7 &   8.00 & 1.500 &   450 &   108 &  20.2 &   129 & F791W &    6.0 &   95.2  & 14 \\ 
NGC\,1140      &  19.8 &   8.28 & 1.024 &    70 &     6 &   6.7 &    51 & F814W &   24.5 &  352.9  & 1, 9, 15 \\ 
NGC\,1156      &   6.5 &   8.39 & 1.630 &    30 &   353 & 240.0 &    87 & F814W &   10.7 &   12.7  & 12, 16 \\ 
NGC\,1741-A    &  54.6 &   8.12 & 1.130 &   120 &    88 &  88.9 &   137 & F814W &    8.5 &  337.5  & 3, 9, 15 \\ 
NGC\,5253-1    &  2.5 &   8.17 & 1.190 &   290 &  1603 & 223.9 &   181 & F814W &    8.3 &    7.6  & 4, 17 \\ 
Pox\,186       &  17.5 &   7.74 & 1.694 &   350 &    36 &   3.5 &   375 & F814W &    4.2 &   53.8  & 18 \\ 
SBS\,0335$-$052  &   53.7 &   7.30 & 2.000 &   412 &    10 &   1.9 &   230 & F791W &    3.5 &  134.4  & 2, 15, 19, 20 \\ 
%SBS\,1159$+$545  &  54.7 &   7.54 & 1.830 &    91 &     5 &   1.3 &   159 & F160W &    3.2 &  170.9 & 15, 21 \\ 
SBS\,1415$+$437  & 12.8 &   7.61 & 1.700 &    60 &    22 &   4.5 &   166 & F791W &    8.0 &   74.7  & 22 \\ 
Tol\,1214$-$277  &  115.6 &   7.55 & 1.979 &   325 &     7 &   1.9 &   290 & F814W &    4.3 &  358.2  & 15, 23 \\ 
Tol\,1924$-$416a &  42.4 &   7.94 & 1.390 &    78 &    64 &   9.7 &    59 & F814W &    2.0 &   61.3  & 24, 25, 26 \\ 
Tol\,65        &  32.0 &   7.54 & 1.732 &   120 &    17 &   2.6 &   221 & F814W &    4.3 &   98.5   & 23, 27 \\ 
UGC\,4483     &   5.1 &   7.58 & 1.700 &   100 &    21 &   2.2 &   169 & F814W &   12.4 &   46.4  & 15, 21\\ 
UM\,311        &   22.6 &   8.31 & 0.970 &    44 &    26 &  10.5 &   279 & F814W &    2.7 &   43.9  & 13, 15 \\ 
% UM\,382        &  46.6 &   7.82 & 1.619 &    45 &     4 &   0.6 &   135 & F160W &    3.9 &  179.8 & 28 \\ 
% UM\,461        &  12.7 &   7.81 & 1.640 &   104 &    23 &  10.3 &   203 & F160W &    2.9 &   35.6 & 8, 13, 15 \\ 
VII\,Zw\,403     &  1.5 &   7.73 & 1.540 &    41 &    45 &   4.7 &   153 & F814W &   12.2 &   13.1   & 10, 12, 15 \\ 
\hline
\end{tabular}
}
\begin{list}{}{}
\item[$^{\mathrm{a}}$] Distance from NED.
\item[$^{\mathrm{b}}$] Oxygen abundance in ionized gas.
\item[$^{\mathrm{c}}$] Electron temperature.
\item[$^{\mathrm{d}}$] Electron number density \nesii\ measured
from the ratio of $\lambda\lambda$6717,6731\,\AA.
\item[$^{\mathrm{e}}$] RMS electron number density \nerms\ inferred
from the H$\beta$ emission measure with diameters as in Col. 11, corrected
for extinction and ionized helium as described in the text.
\item[$^{\mathrm{f}}$] H$\beta$ flux in the spectroscopic
slit.
\item[$^{\mathrm{g}}$] Equivalent width of H$\beta$ emission.
\item[$^{\mathrm{h}}$] Diameters of brightest star-forming
complex as described in the text, without the enlargement factor.
\item[$^{\mathrm{i}}$] Diameters of brightest star-forming
complex (1.5$\times$ that measured in the continuum) as described in the text.
\item[$^{\mathrm{j}}$] Spectroscopic references:
1) Izotov \& Thuan (\cite{izotov04b});
2) Thuan \& Izotov (\cite{thuan05});
3) Vacca \& Conti (\cite{vacca92});
4) Kobulnicky et al.\ (\cite{kobulnicky99});
5) Johnson et al.\ (\cite{johnson00});
6) Kniazev et al.\ (\cite{kniazev00});
7) Corbin et al.\ (\cite{corbin05});
8) Izotov et al.\ (\cite{izotov06});
9) Guseva et al.\ (\cite{guseva00});
10) Izotov et al.\ (\cite{izotov97});
11) Noeske et al.\ (\cite{noeske00});
12) Hunter \& Hoffman (\cite{hunter99});
13) Izotov \& Thuan (\cite{izotov98});
14) Thuan et al.\ (\cite{thuan96});
15) Izotov et al.\ (\cite{izotov07});
16) Moustakas \& Kennicutt (\cite{moustakas06});
17) Walsh \& Roy (\cite{walsh89});
18) Guseva et al.\ (\cite{guseva04});
19) Izotov et al.\ (\cite{izotov99});
20) Izotov et al.\ (\cite{izotov_schaerer06});
21) Izotov et al.\ (\cite{izotov94});
22) Guseva et al.\ (\cite{guseva03});
23) Izotov et al.\ (\cite{izotov_chaffee01});
24) Guseva et al.\ (\cite{guseva07});
25) Papaderos et al.\ (\cite{papaderos06});
26) Kehrig et al.\ (\cite{kehrig06});
27) Izotov et al.\ (\cite{izotov04a});
28) Kniazev et al.\ (\cite{kniazev01}).
\end{list}
  \end{table*}

\appendix

\section{Simple derivation of Eq.\ (\ref{eq:tauSd})}
\label{app:analytic}

Eq.\ (\ref{eq:tauSd}) is derived in Hirashita et al.\
(\cite{hirashita01}) based on Spitzer (\cite{spitzer78}).
Here, we present a simpler derivation of
Eq.\ (\ref{eq:tauSd}).

We start from the optical depth at the threshold
wavelength of the Lyman continuum (912 \AA)
denoted as $\tau_\mathrm{L}$:
\begin{eqnarray}
\tau_\mathrm{L}\simeq A_{1000}/1.086=12.0E_{B-V}\, ,
\end{eqnarray}
where $\tau_\mathrm{L}$ is approximated with the
extinction at $\lambda =1000$ \AA, which is
related with the $B-V$ color excess, $E_{B-V}$, by
assuming the Galactic extinction curve
(Spitzer \cite{spitzer78}). The factor 1.086 is
required for the conversion
from $A_{1000}$ (the extinction at 1000 \AA\ in
units of magnitude) to $\tau_\mathrm{L}$.
Assuming that the dust-to-gas ratio is the same
between the ionized and neutral media
(Sect.\ \ref{subsec:taud}), we simply use
the proportionality between the color
excess and the hydrogen column density derived for
the Galactic environment
(Spitzer \cite{spitzer78}); i.e., we adopt
\begin{eqnarray}
N_\mathrm{H}=5.9\times 10^{21}E_{B-V}~\mathrm{mag}^{-1}~
\mathrm{cm}^{-2}\, ,
\end{eqnarray}
for solar metallicity. Then we obtain
\begin{eqnarray}
\tau_\mathrm{L}=2.03\times 10^{-21}N_\mathrm{H}\, .
\end{eqnarray}

Next, we estimate
$\tau_\mathrm{Sd}$ (the dust optical depth for the ionizing
photons over a path length equal to the Str\"{o}mgren
radius) introduced in Eq.\ (\ref{eq:yi}).
Noting that the hydrogen column density over the
Str\"{o}mgren radius is given by
$N_\mathrm{H}=n_\mathrm{H}r_\mathrm{S}$, we obtain
$\tau_\mathrm{L}=\tau_\mathrm{Sd}$
as shown in Eq.\ (\ref{eq:tauSd}) after using the
expression in Eq.\ (\ref{eq:stromgren}).
If we consider the optical depth over the ionization radius,
$\tau_\mathrm{L}=\tau_\mathrm{di}$ should be applied.

%% reference list

\end{document}